\documentclass{aa}  

\usepackage{graphicx}
\usepackage{txfonts}
\usepackage[]{hyperref}
\usepackage{spverbatim}
\usepackage{multicol}
\begin{document}

   \title{Tidal tails of open clusters}

   \subtitle{}

   \author{Janez Kos
          \inst{1}
          }

   \institute{Faculty of mathematics and physics, University of Ljubljana, Jadranska 19, 1000 Ljubljana, Slovenia\\
              \email{janez.kos@fmf.uni-lj.si}
             }

   \date{Received September 15, 1996; accepted March 16, 1997}

 
  \abstract{Open clusters that emerged from the star forming regions as gravitationally bound structures are subjected to star evaporation, ejection, and tidal forces throughout the rest of their lives. Consequently they form tidal tails that can stretch kiloparsecs along the cluster's orbit.}
   {Cluster members are typically found by searching for overdensities in some parameter space (positions and velocities or sometimes actions and orbital parameters of stars). However, this method is not effective in identifying stars located in the tidal tails far from the open cluster cores. We present a probabilistic method for finding distant cluster members without relying on searching for overdensities and apply it to 476 open clusters.}
   {First, we simulate the dissolution of a cluster and obtain a probability distribution (likelihood) describing where cluster members are to be found. The distribution of stars from the \textit{Gaia} DR3 catalogue in high likelihood regions is then compared to the simulated stellar population of the Galaxy to define the membership probability of each star.}
   {The survey of cluster members included all stars with a magnitude of $G<17.5$ and larger clusters with an age of $>100\ \mathrm{Myr}$ within $3\ \mathrm{kpc}$ from the Sun. We successfully find stars with high membership probabilities in the tidal tails of most clusters, stretching more than a kiloparsec from the cluster cores in some cases. We analyse the morphological properties of tidal tails, demonstrate how properly normalised membership probabilities aid systematic studies of open clusters and publish a catalogue of stars found in the tidal tails.}
  {}

   \keywords{methods: statistical -- surveys -- stars: kinematics and dynamics -- open clusters and associations}

   \maketitle
%
\nolinenumbers
\section{Introduction}
\nolinenumbers
Tidal tails of open clusters were predicted long ago \citep{bok34, spitzer40} \citep[also][and references therein]{kupper08}, but their observational confirmation has only recently been possible with the precise positions and motions of stars obtained by the \textit{Gaia} mission \citep{meingast19, roser19}. Unlike globular clusters and dwarf galaxies, many of which possess well studied tidal tails \citep[e.g.][]{odenkirchen01, belokurov06, ibata19}, open clusters are around a thousand times smaller than globular clusters and are found in the environment around a thousand times more densely populated with stars than the halo. Hence the tidal tails of open clusters do not form a statistically significant overdensity in space. Stars in tidal tails can only be found with algorithms designed for the specific task (unlike general algorithms for finding clusters and streams like OC finder \citep{castroginard18} or STREAMFINDER \citep{malhan18}) and in datasets with small enough uncertainties, so stars in the tail can be delineated from the field stars with some certainty. 

So far, only about a dozen open clusters in the literature have known tidal tails. The first stars in tidal tails well past the cluster tidal radius were found in Hyades. Stars were found in the 6D position-velocity space \citep{meingast19}, as well as in 5D position-proper motion space \citep{roser19}, accounting for the projection effects with the convergent point method \citep{leeuwen09}. Hyades have been extensively researched since then, revealing more stars in the tails stretching up to $800\ \mathrm{pc}$, and showing some structure and overdensities in the tail \citep{jerabkova21}. The shape of the tails and the overdensities are used to infer the best models of cluster evolution in the Galactic gravitational potential and interaction with giant molecular clouds or spiral arms \citep{oh20, jerabkova21, evans22, thomas23}. 

Tidal tails and coronae or haloes of ten open clusters were found by \citet{meingast21} using a convergent-point technique and machine learning. Only a few more clusters have observed tidal tails; M67 \citep{gao20}, NGC 2506 \citep{gao20b}, NGC 752 \citep{boffin22}, IC 4756 \citep{ye21}, and Coma \citet{tang19}. Some more clusters have member stars known outside their core, yet not in extended tidal tails \citep{bhatta22, tarricq21}. Recently, many already disintegrated structures have also been found \citep{kounkel19}, and many of them stretch along distances similar to those expected for the tails of dissolving open clusters \citep[e.g.][]{andrews22}, although most structures have been disputed to represent stars with a common origin \citep{zucker22}.

It is common to all the discoveries of tidal tails in the existing literature that only stars based on high likelihood are found (as opposed to stars with high membership probability). Likelihood is the probability that a star originating from the cluster is found at some coordinate in some parameter space occupied by a cluster. Such selection of most likely cluster members can be reliable \citep{bouma21}, but this is hard to verify without additional observations of radial velocities, asteroseismic parameters, chemistry of stellar atmospheres etc. Membership probability is the probability that a star at some coordinate in some parameter space is a cluster member. Stars with high likelihood do not necessarily have high membership probability, if, for example, they lie in the Galactic plane, where many field stars pass the likelihood threshold as well. To these stars, we can assign some low membership probability, although it would be impossible to know which of the stars are cluster members and which are not -- just that the ensemble of stars has some membership probability. The membership probability would be proportional (or equal, if correctly normalised) to the fraction of the stars that we expect to be cluster members. Not being able to estimate membership probabilities is a strong drawback of existing approaches to finding tidal tails of open clusters. First, there is no objective threshold for the likelihood that delineates members from non-members. This also holds for machine learning techniques, where some internal parameter or a variable dictates the size of found groups. Second, with the maximum likelihood type of methods, it is always possible to find some tidal tails if the underlying model predicts them. This means that there will always be a model for the shape of tidal tails that matches the observations. Hence it is impossible to perform model matching to observed distributions of possible tidal tail stars, if membership probabilities cannot be estimated and the effect of the background population of stars to be evaluated. 

Our work avoids the use of clustering algorithms or machine learning. Instead, we rely on clear and discrete criteria for the selection of possible cluster members, i.e. stars passing some likelihood threshold. This, together with a plain selection function, allows us to apply the same selection criteria on the observed stars and a simulated Galactic population to calculate normalised cluster membership probabilities. With these probabilities, we can analyse the distribution of stars and the shapes of the tidal tails, even in areas far away from the cluster centres where only low-probability members may be present in available data.

The shape and structure of tidal tails are influenced by the cluster formation processes, the gravitational field of the Galaxy, and gravitational perturbations experienced throughout the cluster's lifetime. The positions of the tail \citep{dinnbier22} and epicyclic overdensities \citep{kupper10, jerabkova21} are the easiest to measure, while the prominence of the tail (or length) is the hardest due to uncertainties in membership probabilities far from the cluster core. Observations of tidal tails can shed light on several processes and phenomena discussed below and have been proposed in the literature, some of which are analogous to tidal tails observed in the Galactic halo.

Unlike the tidal tails observed in the Galactic halo, the tidal tails of open clusters can be used to probe the gravitational potential of the inner Galaxy. This offers a unique opportunity to study the inner Galaxy by tracing the kinematical processes like the dissolution of clusters, rather than just observing a static distribution of stars or their orbital parameters. Open clusters in the Solar neighbourhood are affected by the gravitational potential of the Galactic bar, and their orbits offer insights into the bar's rotation and pattern speed. The pattern speed of the bar primarily affects the shape of the tidal tails due to resonances \citep{thomas23}. Clusters offer a unique insight into the rotation of the bar in the last $~2$ Gyr, possibly revealing the pattern speed of the bar slowing down \citep{chiba21}. Due to a wide range of ages and orbits, even open clusters in the Solar neighbourhood offer a unique insight into temporal variations of gravitational potentials involved in tidal tail formation.

In addition to the tidal forces, the clusters are regularly perturbed by the gravitational effects of structures they encounter along their orbits, mostly the giant molecular clouds (GMCs). Such interaction has been explored as the reason for asymmetric tidal tails in Hyades \citep{jerabkova21}. It is expected that such interactions are common and should be a dominant source of perturbation when a cluster passes through the Galactic plane \citep{gieles06, martinez17}. 

The shape of the tidal tails also depends on the gravitational potential of the cluster itself, particularly during the gas expulsion phase immediately after the cluster and star formation \citep{dinnbier20}. Early phases of cluster lives can thus be explored indirectly with a large sample of clusters (to avoid degeneracy between early processes and later perturbation). Cluster rotation \citep{guielherme23} is another internal factor that can affect the shape and prominence of the tidal tails. The evolution of the tidal tails can also be affected by massive objects in the cluster, like OB stars and black holes \citep{wang21}.

Initial mass function for stars, often measured in (young) open clusters, is the critical parameter relating the theories of star formation to observed stellar populations \citep{krumholz19}. Formation of tidal tails can skew the observed mass function, if some stars are not accounted for \citep{gieles09}. Therefore, a more complete census of stars in the tidal tails can contribute to observed mass functions of open clusters to be less affected by mass segregation. This will allow us to infer the initial mass function of older clusters more accurately.

In this paper, we focus on finding stars in tidal tails of open clusters and only briefly discuss the implications of found structures for the study of physical processes involved in their formation. We acknowledge that further research would require more precise n-body simulations and careful matching of tested models to observations. We offer a method of finding normalised probabilities for stars in the tidal tails being members of open clusters. This method is necessary for comparing different models.

This paper first describes the parameter space in which we search for cluster members and tidal tails (Section \ref{sec:space}). Then we present the simulations of the dissolution of open clusters in Section \ref{sec:model}, and in Section \ref{sec:prob} our method for assigning membership probabilities to stars found in the tidal tails. Products and results of our analysis are presented in Section \ref{sec:results}. We also provide online tables with parameters of analysed clusters and stars, as well as figures illustrating the tidal tails of each cluster. In Section \ref{sec:discussion} we compare our results for four clusters with already known tidal tails and discuss some initial findings on the morphology and physics of tidal tails. Discussion on how to use our catalogue is also included in the latter Section. Section \ref{sec:conclusions} finishes with conclusions.

\section{Projected parameter space}
\label{sec:space}

Ideally, the cluster members would be searched for in a parameter space where members form as tight of a group as possible, while field stars are scattered widely. Actions and orbital parameters are often used to construct a parameter space. This works extremely well for finding stellar streams \citep{helmi20} and moving groups \citep{antoja08}, and is in principle a good method for finding open clusters as well \citep{krumholz19}. However, very precise positions and velocities are needed in practice. Clusters in action and orbital parameter space are blurred mostly due to uncertainties in distance and radial velocity measurements. Distances almost always have uncertainties much larger than the size of the cluster, and the radial velocities are not always available and are additionally distorted by binary stars. Therefore a dataset of good enough quality to support finding tidal tails of open clusters in the mentioned parameter spaces would be rather small, limited to brighter and nearby stars, and would have a complicated selection function.

In this work, we aim to find individual stars in the tails of known open clusters. These stars are sparse and a dataset with the emphasis on quantity over quality suits us much better. To derive membership probabilities we also need a simple and well-known selection function. Hence we can only search for cluster members in parameter spaces that can be constructed without accounting for the radial velocity of stars.

We can use the five observables, which include two positions on the sky, two proper motions, and parallax, as the parameter space for finding cluster members in the cluster cores \citep{castroginard18}.  Naturally, stars in the tidal tails do not form a concentrated cluster in this parameter space, but can sometimes be recovered by cluster finding algorithms that search for extended structures \citep{meingast21}. The shape of the clusters in the observables parameter space can be further deformed by projection effects. These can be untangled with the convergent point method \citep{debruijne99, roser19} to find members further away from the cluster cores. The general structure of tidal tails can also be accounted for (if known from n-body simulations for example) by a so-called compact convergent point algorithm \citep{jerabkova21}. 

Similarly to the convergent point methods, we search for cluster members in a transformed parameter space. Using \texttt{Astropy}'s \texttt{SkyOffsetFrame} function that perform coordinate and velocity conversions via Euler angles \citep[e.g.][]{tatum}, we can calculate the distance of a star from the cluster centre, the position of the star in a coordinate system with the cluster in the origin, and the velocities in this coordinate system, i.e. the velocity of the star in the radial direction from the cluster centre ($v_\parallel$) and in the orthogonal direction ($v_\perp$), defined as
\begin{equation}
    v_{\parallel}=\frac{\mathbf{\Delta\mu}\cdot\mathbf{\Delta c}}{|\mathbf{\Delta c}|},
\end{equation}
and
\begin{equation}
    v_{\perp}=\sqrt{\left(\mathbf{\Delta\mu}\right)^2-\left(\frac{\mathbf{\Delta\mu}\cdot\mathbf{\Delta c}}{|\mathbf{\Delta c}|}\right)^2},
\end{equation}
where vector $\mathbf{\Delta\mu}$ contains proper motions translated into the system where the cluster is at the origin, and $\mathbf{\Delta c}$ is a vector of celestial coordinates in the same translated system. 

It is obvious that in the absence of projection effects and long tidal tails, the stars that were ejected from the cluster must have $v_\parallel$ proportional to the distance from the cluster ($d$), and $v_\perp$ must be close to zero. Projection effects cause the relation between $d$, $v_\parallel$, and $v_\perp$ to bend away from a straight line, and forces shaping the tidal tails can make the relationship between the above quantities even more complicated. However, the stars in the cluster, or originating in the cluster, will always form a coherent and continuous shape in the projected parameter space if the perturbation comes only from tidal forces. The stars will also occupy a smaller region of the projected parameter space (form a tighter group) than in the observed proper motion space. The projected parameter space is illustrated in Section \ref{sec:nonprob} for an example of NGC 2516.

\section{Modelling dissolution of clusters}
\label{sec:model}

The first step in our algorithm for finding distant cluster members is defining a likelihood that describes where in a parameter space we expect to find cluster members. This is done by a simulation, where we take the positional and kinematic parameters of a known cluster, integrate the orbit of the cluster centre back in time, populate a cluster with stars and then do a simulation of the cluster evolution forward in time to the present age. The result is a simulated cluster that samples the wanted likelihood. It is essential that the likelihood is sampled with sufficient density, which in our case, where cluster members are searched for in a 5D parameter space, means up to one million samples per cluster, depending on the size of the tidal tails. Note that the positions of simulated stars in each dimension are well correlated, so simulating only $\sim 15$ stars per dimension still produces enough samples.

\subsection{Initial conditions}

For clusters younger than $750\ \mathrm{Myr}$ we simulate their dissolution throughout their whole life. At most, this takes roughly three revolutions around the Galaxy.  In simulations, the clusters with the age of $750\ \mathrm{Myr}$ develop tidal tails the length of a few $\mathrm{kpc}$. It is extremely unlikely that we would be able to find any cluster members at these distances either due to error propagation, a large size of the parameter space, or a small number of members that should exist at these distances in the first place. We therefore only simulate the last $750\ \mathrm{Myr}$ of dissolution for older clusters. For longer simulations, we would have to account for different Galactic gravitational potentials, as small variations can have a significant impact on the orbit after a while. This would produce a spread-out likelihood distribution that favours many stars as members, but all with tiny probabilities. We do not have any mechanism at hand that would reduce such a large amount of low probable members into a useful or applicable sample. Even the simulation of the last $750\ \mathrm{Myr}$ can be significantly uncertain for the shape of the orbits $>1\ \mathrm{kpc}$ from the cluster centre. 

\subsubsection{Galactic potential}

\begin{figure}
    \centering
    \includegraphics[width=\columnwidth]{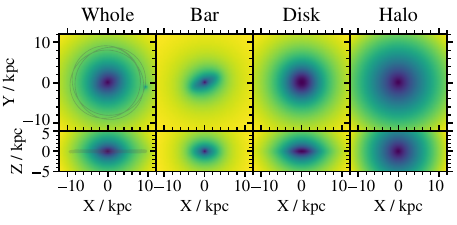}
    \caption{Illustration of Galactic potentials used in the simulation of stellar orbits. The left column shows the complete gravitational potential, and the following columns show the potential of the bar, disk and the halo. The black line shows the orbit of the Sun for the last 750 Myr. A small dip close to the present position of the Sun is the gravitational potential of a simulated cluster, amplified one million times. Magnitudes of different potentials are not plotted to scale.}
    \label{fig:potentials}
\end{figure}

To simulate the orbits of clusters we constructed a 4 component Galactic potential. The halo is described by a Navarro-Frenk-White potential (scale radius $R=16.244\ \mathrm{kpc}$, normalised so its rotation curve peaks at $r_\mathrm{max}=35.13\ \mathrm{kpc}$ with $v_\mathrm{max}=169.0\ \mathrm{km\,s^{-1}}$), a bar with a softened needle bar potential (mass $m=10^{10}\ \mathrm{M_\odot}$, bar half length $a=3.5\ \mathrm{kpc}$, prolate softening length $c=1.0\ \mathrm{kpc}$, pattern speed $\Omega_b=1.85$, and position angle at present time $\phi_\mathrm{PA}=0.4$), bulge with a Plummer potential (mass $m=2.0\,10^{9}\ \mathrm{M_\odot}$, scale $c=0.25\ \mathrm{kpc}$), and the disk is described by a Miyamoto Nagai potential (mass $m=6.0\,10^{10}\ \mathrm{M_\odot}$, scale radius $R=3.0\ \mathrm{kpc}$, scale height $h=0.28\ \mathrm{kpc}$). Figure \ref{fig:potentials} illustrates the gravitational potential used in this work. The code to produce the potential is given in Appendix \ref{sec:galpy}.

Potentials, distances and velocities are normalised or converted to natural units using $R_\odot=8.122\ \mathrm{kpc}$ for the Sun's distance to the Galactic centre \citep{gravity18} and $v_\odot=233.4\ \mathrm{km\,s^{-1}}$ for the circular velocity of the LSR \citep{drimmel18}, i.e. the values used by default in \textsc{Astropy} for coordinate system conversions. Orbits are then integrated using \textsc{galpy} package. 

To find the initial conditions for the cluster centre, we integrate its centre back in time using the potential mentioned above. For the actual simulation of cluster's dissolution, we add the gravitational potential of the cluster and populate it with massless particles.

\subsubsection{Clusters parameters}
\label{sec:cluster_params}
To simulate a cluster we use the position and velocity of a cluster at the present time and integrate back the orbit of the cluster's centre. There we initialise the cluster and simulate the kinematics of its stars. The positions, proper motions, distances, and ages (also used to generate an isochrone and calculate stellar masses) of clusters are taken from \citet{cantatgaudin20}. We analyse all clusters from this catalogue with age $>100\ \mathrm{Myr}$, distance $<3\ \mathrm{kpc}$, with more than 45 stars found by \citet{cantatgaudin20} (parameter \texttt{nbstars07} $> 45$) and with available radial velocities (see below). Open clusters analysed in this study therefore exclude any poorly defined structures, large dissolved groups or associations. Additionally to 475 clusters that satisfy the above conditions, we analysed Pleiades (Melotte 22). Pleiades have an age of $77.6\ \mathrm{Myr}$ in \citet{cantatgaudin20}. However, in the literature their age is usually $100\ \mathrm{Myr}$ or more \citep[e.g.][]{gossage18, bouvier18, murphy22, hunt23}. Because Pleiades is a commonly studied cluster, also in the context of tidal tails \citep{meingast21}, we include Pleiades in our analysis. Age and parameters from \citet{cantatgaudin20} were used in simulations and analysis for Pleiades, same as for all other clusters.

We take radial velocities from Simbad. Clusters in the above catalogue with no radial velocity in Simbad were not analysed. The source of the radial velocity is given in the online table of cluster parameters (\textit{Clusters} table). Radial velocity sources are \citet{conrad14, conrad17, loktin17, gaia18, kos18, soubiran18, carrera19, casali19, dias19, monteiro19, donor20, zhong20, dias21,  magrini21, tarricq21, carrera22}. For 43 clusters we repeated the simulation with our radial velocities obtained from stars initially found as members with the Simbad's radial velocity (see Section \ref{sec:newrv}).

Coordinate conversion between the celestial and Galactic coordinate systems was done by \textsc{Astropy} using $(U,\  V,\  W)_\odot=(-12.9,\ 245.6,\ 7.78)\ \mathrm{km\,s^{-1}}$ \citep{drimmel18}, $R_\odot=8.122\ \mathrm{kpc}$ \citep{gravity18}, and $z_\odot=20.8\ \mathrm{pc}$ \citep{bennett19}. 


\subsubsection{Cluster initialisation}

Because we are not interested in the dynamical properties of stars inside the cluster, but rather just the stars very far away from the cluster core, we did not intend to simulate the core itself. Kinematics in the core are dominated by multi-body interactions, which would require a precise and computationally slow simulation, possibly involving stellar evolution as well. Motions of stars far away from the cluster core are completely dominated by the Galactic gravitational potential. Once the star is ejected or stripped from the cluster, the intra-cluster dynamics are irrelevant to the star's future orbit. Therefore we only aim to simulate well the ejection of stars from cluster core, tidal evaporation and the subsequent orbit of a star in the Galactic potential. We also neglect any gravitational effects of spiral arms or giant molecular clouds, as their evolution throughout the last $750\ \mathrm{Myr}$ is mostly unknown. 

All clusters are initialised in the same way, regardless of their age, mass or star counts; we create a Plummer potential well centred at the cluster with a scale parameter (radius) of $2\ \mathrm{pc}$ and a mass of $5000\ M_\odot$. We populate this potential with one million point masses, following the recipe in \citet{aarseth74}, so they form a stable distribution that would give rise to the original Plummer potential. Masses of bodies are irrelevant, as our simulation has no body-body interactions. 

\subsubsection{Ejection and tidal evaporation}
\label{sec:escape}

Stars are removed from a cluster by two mechanisms: tidal evaporation and two/three-body interactions. We simulate the former, but not the latter. Tidal evaporation works on edges of the cluster close to the cluster -- Galactic centre line (Lagrange points, \citep{fukushige00}) on long timescales. Because the initial conditions for a cluster are stable (the cluster is gravitationally bound), we expect very little tidal evaporation. Because a gravitationally bound cluster is not the most realistic representation of known open clusters, and to move more stars into tidal tails, we initialise a mass loss from the Plummer potential. All clusters start with a mass of $5000\ M_\odot$ and lose $4000\ M_\odot$ by the present time. Note that such simplification does not simulate the velocity distribution in the cluster core well. 

Two- or three-body interactions are more common at the beginning of the cluster's life and are random in terms of direction. They cause mass segregation, mix the initial velocity distribution and cause fast ejections of stars. We do not simulate these processes directly, but instead assume and analytically model a velocity distribution and rate for ejected stars. These stars will be the only ones that can end up far away from the cluster core of young clusters and allow us to sample the distribution of stars in the halo or corona of the cluster. 

The ejection velocities are random in direction, but the amplitude changes with time as:
\begin{equation}
    v_\mathrm{eject}=\left[300.0\left(\frac{t}{\mathrm{Myr}}\right)^{-1.7}+1.0\right]\ \mathrm{km\ s^{-1}}.
\end{equation}
This function approximates the results of a simulation in \citet{moyano13}. The rate of ejected stars changes with time as
\begin{equation}
    P_\mathrm{e}(t)=-\frac{1}{3}\frac{t}{\tau}+\frac{2}{3},
\end{equation}
where $P_\mathrm{e}(t)$ is the probability that a star is ejected at time $t$, and $\tau$ is the age of the cluster. Half of the stars in the cluster will get the kick described above once in the simulation. Again, this is not realistic, but sends more stars into the tidal tails and the cluster's halo, where we want to sample the likelihood as well as possible. 

For each star we calculate when and how it should be ejected, and later simulate it in two parts; until the ejection, after which we add the kick, and then simulate the rest of the orbit. 

\subsection{Orbit integration}
\label{sec:orbits}

When simulated stars are initialised, we integrate their orbits in the combined gravitational potential of the Galaxy and the cluster, where the gravitational potential of the cluster is moving along the pre-calculated orbit and is getting shallower with time, as described in the previous section.

Orbits are integrated with a Runge-Kutta type of algorithm with a relative precision of $10^{-3}$ and absolute precision of $10^{-6}\ R_\odot$. This is sufficient, so no energy is lost while simulating the dynamics of tidal tails. Such simulation is fast enough that it can be performed for one million bodies in hundreds of clusters in a reasonable time on one node of a cluster computer. 

\subsection{Constructing the membership likelihood distribution}
\label{sec:unc}

The result of a cluster dissolution simulation is a sampled probability distribution for the present location of cluster members. We call this probability distribution membership likelihood. It tells us where we expect to find cluster members, but not yet the membership probability of a found star, as the likelihood does not account for any underlying structure of the Galaxy and distribution of field stars. 

\textit{Gaia} stars for which we want to calculate membership likelihoods are affected by observational uncertainties. Ideally, the likelihood for each star would be calculated by multiplying the membership likelihood distribution with the distribution describing the measured parameters of a star. This is computationally too expensive, so instead we convolve the simulated likelihood distribution with \textit{Gaia}'s mean observational uncertainties \citep{fabricius21, katz23}. The simulated likelihood is re-sampled with a multivariate Gaussian kernel using the following uncertainties (multiple values indicate a sum of multiple Gaussian kernels with given widths):
\begin{multline}
    \langle\sigma_{(\alpha\, \delta)}\rangle = 0.04\ \mathrm{mas}\\
    \langle\sigma_\varpi\rangle = 0.03,\ 0.07\ \mathrm{mas}\\
    \langle\sigma_{(\mu_\alpha\, \mu_\delta)}\rangle = 0.03,\ 0.08\ \mathrm{mas\, yr^{-1}}\\
    \langle\sigma_{v_r}\rangle = 0.5,\ 2.0,\ 6.0, 150, 300\ \mathrm{km\, s^{-1}.}\hfill
\end{multline}
Only $9\%$ of samples are convolved with a $ v_r=150\ \mathrm{km\, s^{-1}} $ uncertainty and only $1\%$ with a $v_r=300\ \mathrm{km\, s^{-1}}$ uncertainty, both of which represent unresolved binaries. The rest of the samples are evenly divided between multiple uncertainties listed above. The uncertainties represent well the \textit{Gaia}'s observational uncertainties for stars up to magnitude $G<17.5$. Simulated stars that sample the likelihood have no mass or magnitude attributed to them, so all can be re-sampled using the same mean uncertainties. 

\subsection{Defining likelihoods for \textit{Gaia} stars}
\label{sec:likelihood}

To calculate the likelihood of each \textit{Gaia} star from the sampled likelihood distribution we first have to parameterise the sampled likelihood distribution. We use a parameter space of three positions and two projected velocities, so we need to define a joined five-dimensional parameterisation. This is done by making a five-dimensional grid encompassing the bulk of the distribution of simulated stars and calculating a histogram of star counts on this grid. The number of samples in each bin is proportional to the likelihood ($\mathcal{L}$) value in that bin:
\begin{equation}
    \mathcal{L}\propto N_\mathrm{samples}.
\end{equation}
For each \textit{Gaia} star we find its corresponding bin as well and assign it the likelihood of that bin. All \textit{Gaia} stars in a single bin therefore have the same membership likelihood. Likelihood is reported in the \textit{Members} table under the keyword \texttt{likelihood}. There is no need for any normalisation of likelihoods, as we only use likelihood for relative comparisons between \textit{Gaia} stars being more or less likely the members of one cluster. 

The five-dimensional space is binned with constant steps of $20\ \mathrm{pc}$ in $X$ and $Y$, $15\ \mathrm{pc}$ in $Z$, and $1\ \mathrm{mas\,yr^{-1}}$ in $v_\parallel$ and $v_\perp$. For a small number of clusters, we used proper motions instead of projected proper motions, which were binned in steps of $1.5\ \mathrm{mas\, yr^{-1}}$.

For stars with existing radial velocities, we calculate another likelihood, where $v_\parallel$ and $v_\perp$ are calculated from three cartesian velocities, which are calculated from proper motions and the radial velocity (i.e. from the six-dimensional position and velocity vector). This likelihood is reported in the table \textit{Members} under the keyword \texttt{likelihood\_6d}. Note that this likelihood is still calculated in five dimensions because the simulation would have to be much larger to fill the six-dimensional space compared to a five-dimensional space. 6D likelihood was calculated with the same binning, we just updated $v_\parallel$ and $v_\perp$ with values incorporating the radial velocities.

\section{Assigning cluster membership probabilities}
\label{sec:prob}

Membership likelihood defines a region in the parameter space where stars originating from the cluster are expected to be found. A probability that an actual star is indeed a cluster member depends on the likelihood at the position of the star and the probability of finding a star at that position in the population of field stars alone. In this section we calculate likelihoods of individual stars in the \textit{Gaia} DR3 catalogue and combine them with a model of the Galactic stellar population to derive cluster membership probabilities.

\subsection{Data}

\textit{Gaia} data release 3 data \citep{gaia16, gaia23} was downloaded from the \texttt{gaia\_source} table for all stars with $G<17.5$ in some regions around the cluster and its tidal tails. An example of the query is given in Appendix \ref{sec:query}. No quality cuts were made initially, as low-quality or flagged data represents only a small fraction of the stars, but subjective quality cuts can impact the selection function. Completeness is an important issue in our algorithm, so all quality cuts were made only after the members were found, if necessary. 

We split the Galaxy into regions $18\times9\ \mathrm{deg^2}$ large on the sky and further split into 30 distance bins to the distance of $5\ \mathrm{kpc}$. For each cluster, we only downloaded regions with more than 50 simulated stars. Each region has a low enough number of \textit{Gaia} stars in it so the query on the \textit{Gaia}'s server can be performed and all the provided data can be downloaded. At the same time, we refrain from downloading regions where we are unlikely to find cluster members to save time and computer memory. For each cluster, we end up with an order of one million stars from which we will select cluster members. In rare cases (clusters with very long tails and clusters in the direction toward the inner parts of the Galaxy), the number of stars is larger than 10 million. 

To calculate distances from parallaxes we used a direct inversion without any priors. The first reason is that geometric or photometric priors do not account for congregates of cluster stars \citep{bailer21}. The second reason is that the simulations of the Galactic stellar population are made with distance as the parameter, but \textit{Gaia} queries are made in respect to the parallax. We solve the discrepancy in possible star counts by calculating the parallax of simulated field stars using the uncertainties given in Section \ref{sec:unc} and simulate an appropriately larger region.

\subsection{Non-probabilistic method for finding cluster members}
\label{sec:nonprob}

In Section \ref{sec:space} we described how a parameter space of projected velocities is constructed. In this section, we show an example of the parameter space and illustrate the basic principles of using projected velocities by finding cluster members in NGC 2516 using simple selection criteria in the parameter space. The example also demonstrates the limitations of a non-probabilistic approach to finding cluster members far away from the cluster cores. 

The non-probabilistic approach demonstrated here is similar to the traditional approach, where we search for cluster members in some parameter space (like ($\alpha$, $\delta$, $\mu_\alpha$, $\mu_\delta$, $\varpi$) or any similar combination of coordinates) by selecting the region around the cluster, which then delineates cluster members from non-members. The advantage of this approach is speed, clear selection function, and repeatability. Such a method is incapable of finding very distant cluster members with high confidence if we try to expand the boundaries defining the cluster. It is also impossible to define normalised membership probabilities for found stars without some further effort. We must note, that such a method works much better when all three positions and all three velocities of the stars are known, as one can then search for cluster members in a cartesian parameter space unaffected by projection effects.

\begin{figure*}
    \centering
    \includegraphics[width=\textwidth]{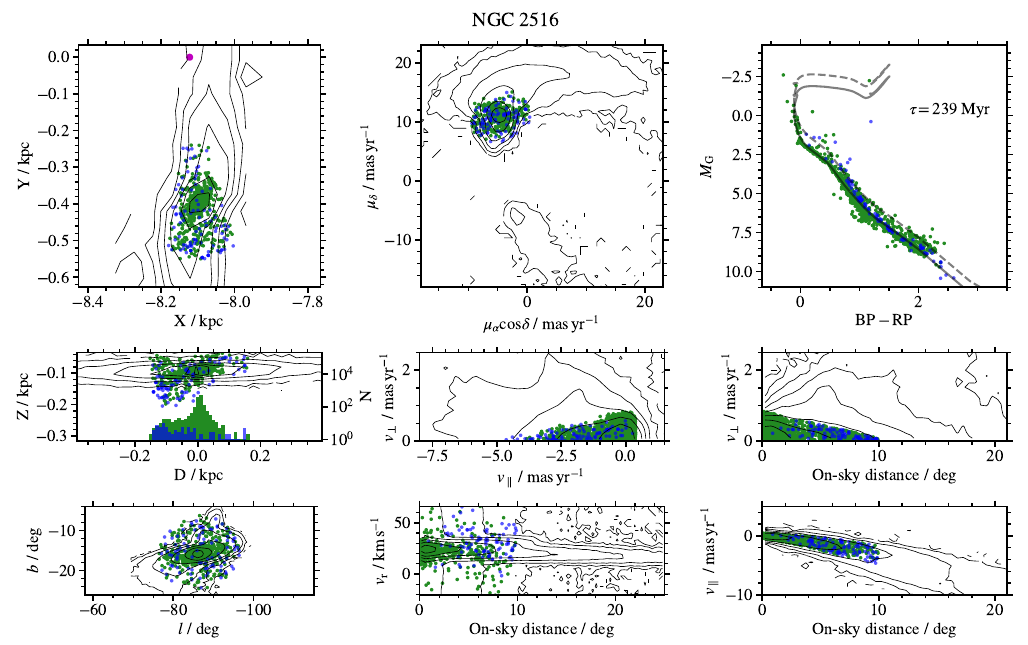}
    \caption{Members of NGC~2516 found with a simple non-probabilistic method (green) shown in the parameter space explored in this work. Simulated field stars found with the same selection as used for finding the members are shown in blue. Black contours show the likelihood for the position of cluster members predicted by a simulation of a dissolving open cluster. Contours are log-spaced. Top row: XY Galactic plane, proper motions space, colour-magnitude diagram. Middle row: Position along the main axis of the tidal tails ($D$) and density of stars along this line, two projected proper motions, $v_\perp$ projected motion as a function of distance from the cluster. Bottom row: Position on the sky, radial velocity as a function of distance from the cluster, and $v_\parallel$ projected motion as a function of distance from the cluster.}
    \label{fig:nonprob}
\end{figure*}

From the simulation of the cluster dissolution (shown with contours in Figure \ref{fig:nonprob}) we can deduce the projection effects: NGC 2516 is receding from us fast enough that it appears to be shrinking when motions of the stars are projected onto the sky. This is evident from the $v_\parallel$ vs. distance panel in Figure \ref{fig:nonprob}. We can select the cluster members following the shape predicted by the simulation. We did the same for $v_\perp$, and then further constrained the extent of the cluster on the sky (radius $d_\mathrm{2D}$) and the range of the parallax to find members of NGC 2516. The complete selection criterion is
    \begin{multline}
        d_\mathrm{2D}<10^\circ,\\
        1.8 < \varpi<4.0\ \mathrm{mas},\\
        v_\parallel > -0.2\, d_\mathrm{2D} -0.015\, d_{\mathrm{2D}}^2 - 0.4 - 0.14\, d_\mathrm{2D}\ \mathrm{mas\,yr^{-1}},\\
        v_\parallel < -0.2\, d_\mathrm{2D} -0.015\, d_{\mathrm{2D}}^2 + 0.4 + 0.14\, d_\mathrm{2D}\ \mathrm{mas\,yr^{-1}},\\
        v_\perp < 0.85 - 0.077\, d_\mathrm{2D}\ \mathrm{mas\,yr^{-1}},\hfill
    \end{multline}
where $d_\mathrm{2D}$ is given in degrees. Members found by the non-probabilistic method in Figure \ref{fig:nonprob} are shown in green. We applied the same selection on a simulated population of stars, that does not include any clusters (see Section \ref{sec:galaxia}). These stars are shown in blue. From Figure \ref{fig:nonprob} we can learn:
\begin{itemize}
    \item Projection effects are complicated in the proper motions plane (the contours originate from the core of the cluster and wrap outside the panel into the convergent point at $(\mu_\alpha,\ \mu_\delta)=(2, -5)\ \mathrm{mas\, yr^{-1}}$). The projection effects are simpler, mostly monotonic in the projected velocity space.
    \item Cluster members occupy a smaller portion of the projected velocity space compared to the proper motion space, i.e. they are better clustered.
    \item Members found furthest from the cluster core are not reliable. This is best illustrated by the histogram in the second panel on the left, where in some regions we find almost as many simulated stars as real stars. Only a small fraction of those stars can be members.
    \item Using the radial velocities would let us make a better selection by removing stars that have radial velocity measured but do not follow the predicted distribution of radial velocities. 
    \item Same could be done using the photometric information and removing the stars that do not follow the isochrone. Note that some scatter around the isochrone is normal due to differential reddening. 
\end{itemize}

Our analysis of NGC 2516 shows results similar to \citet{tarricq22} -- we find an oblong distribution of stars around the core and some halo members, but we are unable to find longer tidal tails like \citet{meingast21}. To use the full information of the simulated likelihood and simulated Galactic population of stars we use probabilistic methods described in the next section. Then we are able to find cluster members with high certainty even in the region where field stars dominate the selection of stars in the above example.

\subsection{\textsc{Galaxia} simulations}
\label{sec:galaxia}

To generate a synthetic survey of the Galaxy we use the \textsc{Galaxia} code \citep{sharma11} and populate the exact same space that was queried in \textit{Gaia} with a synthetic population of stars. The Besancon analytical model of the Galaxy \citep{robin03} that is used is smooth, so there are no clusters in the synthetic population. Note that the stars representing the age, chemical composition and kinematics of the clusters are in the model. They are just not put into clusters but still exist in the general region of the synthetic Galaxy. \textsc{Galaxia} uses PADOVA isochrones \citep{bressan12} and can provide magnitudes of synthetic stars in \textit{Gaia} DR3 passbands \citep{riello21}. The latter is important for the selection function of the simulated stars to be as well matched as possible to \textit{Gaia}'s selection function. 

To avoid low-number statistics in some regions of the simulated space guiding the calculation of membership probabilities, we also oversample the \textsc{Galaxia} simulation by a factor of five. The accuracy of calculated membership probabilities benefits greatly from an oversampled simulation, although an even higher oversampling was not feasible due to computer memory limitations.

\begin{figure}
    \centering
    \includegraphics[width=\columnwidth]{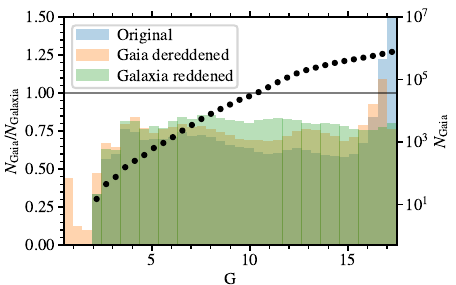}
    \caption{Completeness of the \textsc{Galaxia} simulation vs. the \textit{Gaia}'s. Histograms show the number of stars per magnitude bin in the region of NGC 2516. Blue shows the distribution when no reddening correction is done to the apparent magnitudes. Orange is the distribution when \textit{Gaia}'s magnitudes are corrected for \textit{Gaia}'s extinction, and Green when \textsc{Galaxia} absolute magnitudes are converted to apparent magnitudes with \textsc{Galaxia}'s extinction taken into account. We use the latter magnitudes in this study. Points together with the axis on the right show the number of stars in \textit{Gaia} per magnitude bin downloaded in the region of this cluster.}
    \label{fig:completeness}
\end{figure}

Absolute magnitudes, distance, and interstellar reddening are given by \textsc{Galaxia} for simulated stars. For the simulated population to be comparable to the observed one, we have to correct the magnitudes for the extinction in selected bands ourselves. We used bandpasses from \citet{riello21} to convert \textsc{Galaxia}'s absolute magnitudes to apparent magnitudes using the extinction law with $R_V=3.15$ and calibration factors from \citet{sharma11}. While \textit{Gaia} has extinction measured for most stars, we find much more compatible star counts between \textit{Gaia} and \textsc{Galaxia} if we correct the latter. This is illustrated in Figure \ref{fig:completeness} where completeness is compared when no extinction correction is done (blue histograms), when \textit{Gaia}'s magnitudes are corrected for \textit{Gaia}'s extinction (orange histogram), and when \textsc{Galaxia}'s magnitudes are corrected for extinction modelled by \textsc{Galaxia} (green histogram). The figure shows the case for one cluster, but correcting \textsc{Galaxia}'s magnitudes is always preferred. 

An ideal simulation of synthetic stars would match the star counts by \textit{Gaia} in some broad region around the cluster. We find that this is rarely the case and the star counts can vary by up to $20\%$, see Figure \ref{fig:completeness}. \textit{Gaia} is, for all practical reasons, complete between magnitudes $G=12$ and $17$. Incompleteness is negligible for our case at $G=17.5$ \citep{fabricius21}, the limiting magnitude of this study, as well as at the bright end. Hence the discrepancy between \textit{Gaia} and \textsc{Galaxia} star counts is due to the simulation not encasing the complexity of the real Galaxy. We take this incompleteness into account when we later calculate membership probabilities by scaling all \textsc{Galaxia} counts by the incompleteness value. 

The likelihood of \textsc{Galaxia} stars being members of a cluster are calculated in the same way as the likelihoods for \textit{Gaia} stars in Section \ref{sec:likelihood}. Consequently, all \textit{Gaia} or \textsc{Galaxia} stars have the same likelihood, if they are found in the same bin of the 5D parameter space.

\subsection{Calculating cluster membership probabilities}

We aim to assign a membership probability to all stars in \textit{Gaia} DR3 catalogue that pass a threshold of likelihood for being members of one of the studied clusters. The likelihoods of stars in each cluster have discrete values, as the likelihoods are proportional to the number of simulated stars in each 5D histogram bin. We selected the likelihood threshold so no more than 5000 Gaia stars make the  selection, and then lower the threshold by one discrete step. This cut in likelihoods was done once for each cluster before the star counts described in the next sections were made.

\subsubsection{Definitions}
\label{sec:def}

In Section \ref{sec:likelihood} we described how to assign membership likelihoods to \textit{Gaia} stars. Based on the likelihood, we can select candidates for cluster members. Within a region of the parameter space, the probability that a star is indeed a cluster member equals
\begin{equation}
    P=\frac{N_\mathrm{cluster}}{N_\mathrm{field}+N_\mathrm{cluster}},
    \label{eq:prob}
\end{equation}
where $N_\mathrm{cluster}$ is the number of cluster stars in that region, and $N_\mathrm{field}$ is the number of field stars (all other stars that are not members of the particular cluster) in the same region of the parameter space. We call $P$ the membership probability. Of course, the $N_\mathrm{cluster}$ and $N_\mathrm{field}$ are not known. What we do know is the observed number of stars (called $N_\mathrm{Gaia}$), which is the sum of stars in the cluster and the field stars:
\begin{equation}
    N_\mathrm{Gaia}=N_\mathrm{cluster}+N_\mathrm{field}.
    \label{eq:n_gaia}
\end{equation}
$N_\mathrm{cluster}$ and $N_\mathrm{field}$ cannot be measured, so we assume that $N_\mathrm{field}$ equals the number of simulated stars $N_\mathrm{Galaxia}$. Subsequently, we can derive $N_\mathrm{cluster}$ and calculate the membership probability from Equation \ref{eq:prob}. However, one must be careful, as all numbers above are in reality just one sample taken from a Poisson distribution. The consequence is, that even if $N_\mathrm{Gaia}=N_\mathrm{Galaxia}$, for example, there is some chance that the \textit{Gaia} stars are members of the cluster and we just got unlucky with the sampling of the distribution. 

\begin{figure}
    \centering
    \includegraphics[width=\columnwidth]{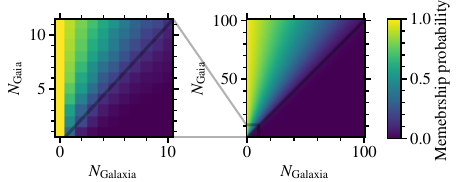}
    \caption{Membership probability lookup table given the number of observed stars ($N_\mathrm{Gaia}$) and the number of simulated stars ($N_\mathrm{Galaxia}$) in any region of the parameter space. The left panel shows a detail of the right panel for a small number of observed or simulated stars.}
    \label{fig:prob_table}
\end{figure}

The probability that a star is a cluster member can be calculated for any combination of  $N_\mathrm{Gaia}$ and $N_\mathrm{Galaxia}$. Formally this involves operating with cumulative distributions of the Poisson distribution, so instead we simulated the membership probability for any combination of $N_\mathrm{Gaia}$ and $N_\mathrm{Galaxia}$. It is evident from Equation \ref{eq:n_gaia} that the same $N_\mathrm{Gaia}$ can be obtained from different combinations of $N_\mathrm{cluster}$ and $N_\mathrm{field}$. We follow Equation \ref{eq:prob} to calculate the probabilities for all combinations of $N_\mathrm{cluster}$ and $N_\mathrm{Galaxia}$, and save them into a tableau. We then convert $N_\mathrm{cluster}$ and $N_\mathrm{Galaxia}$ to $N_\mathrm{Gaia}$ and average probabilities from the tableau for the same combination of $N_\mathrm{Gaia}$ and $N_\mathrm{Galaxia}$. The resulting table is illustrated in Figure \ref{fig:prob_table}. We also account for the oversampling of the \textsc{Galaxia} simulation when calculating the probabilities. We query a lookup table for each bin of the parameter space, which is faster than calculating the probabilities from an equation.

\subsubsection{Integrated and binned membership probabilities}
\label{sec:intbin}

There are several ways of binning the parameter space to get the numbers for $N_\mathrm{Gaia}$ and $N_\mathrm{Galaxia}$ and consequently the membership probabilities for stars in that bin. Ideally, all methods would give the same probabilities, but in reality, this is not true either due to the simulation lacking small scale structure, or due to the small number statistics. We report two different membership probabilities called integrated probability and binned probability (called \texttt{probability\_int} and \texttt{probability\_bin} in the \textit{Members} table). 

To calculated the integrated probability, we use the same bins as defined in Section \ref{sec:likelihood} for calculating likelihoods of stars. We then iterate through the likelihoods in log steps of  $\log \mathcal{L}=0.0385$ and count $N_\mathrm{Gaia}$ and $N_\mathrm{Galaxia}$ stars with likelihood between $\log\mathcal{L}_\mathrm{i}<\log\mathcal{L}<\log\mathcal{L}_\mathrm{i}+0.0385$. Thus we add the star counts from all 5D histogram bins that correspond to the same range of likelihoods to obtain $N_\mathrm{Gaia}$ and $N_\mathrm{Galaxia}$. We than follow Equation \ref{eq:prob} an derivation in Section \ref{sec:def} to calculate membership probabilities from star-counts $N_\mathrm{Gaia}$ and $N_\mathrm{Galaxia}$.

To calculate the binned membership probability, we count $N_\mathrm{Gaia}$ and $N_\mathrm{Galaxia}$ in each individual 5D bin, regardless the likelihood. Because the number of \textit{Gaia} or \textsc{Galaxia} stars in each bin is almost always low (orders of magnitude lower than the number of simulated stars used to calculate likelihoods), we re-bin the data  with bins twice the size (32 times the volume) of the bins used for likelihood calculations. Even after re-binning, the most common number of stars in each bin is zero, and it is also most likely that \textit{Gaia} and \textsc{Galaxia} stars are distributed sparsely with only one of either type of stars in one bin. Hence we re-bin the space further by merging the neighbouring bins, until there is at least one \textit{Gaia} star in each bin, and there are no empty bins. Then we count the number of \textit{Gaia} and \textsc{Galaxia} stars in each bin and calculate the membership probability for stars in that bin. The re-binned space is essentially a discrete Voronoi tessellation in five dimensions with \textit{Gaia} stars defining the cell centres. 

\section{Results}
\label{sec:results}

\begin{figure*}
    \centering
    \includegraphics[width=0.98\textwidth]{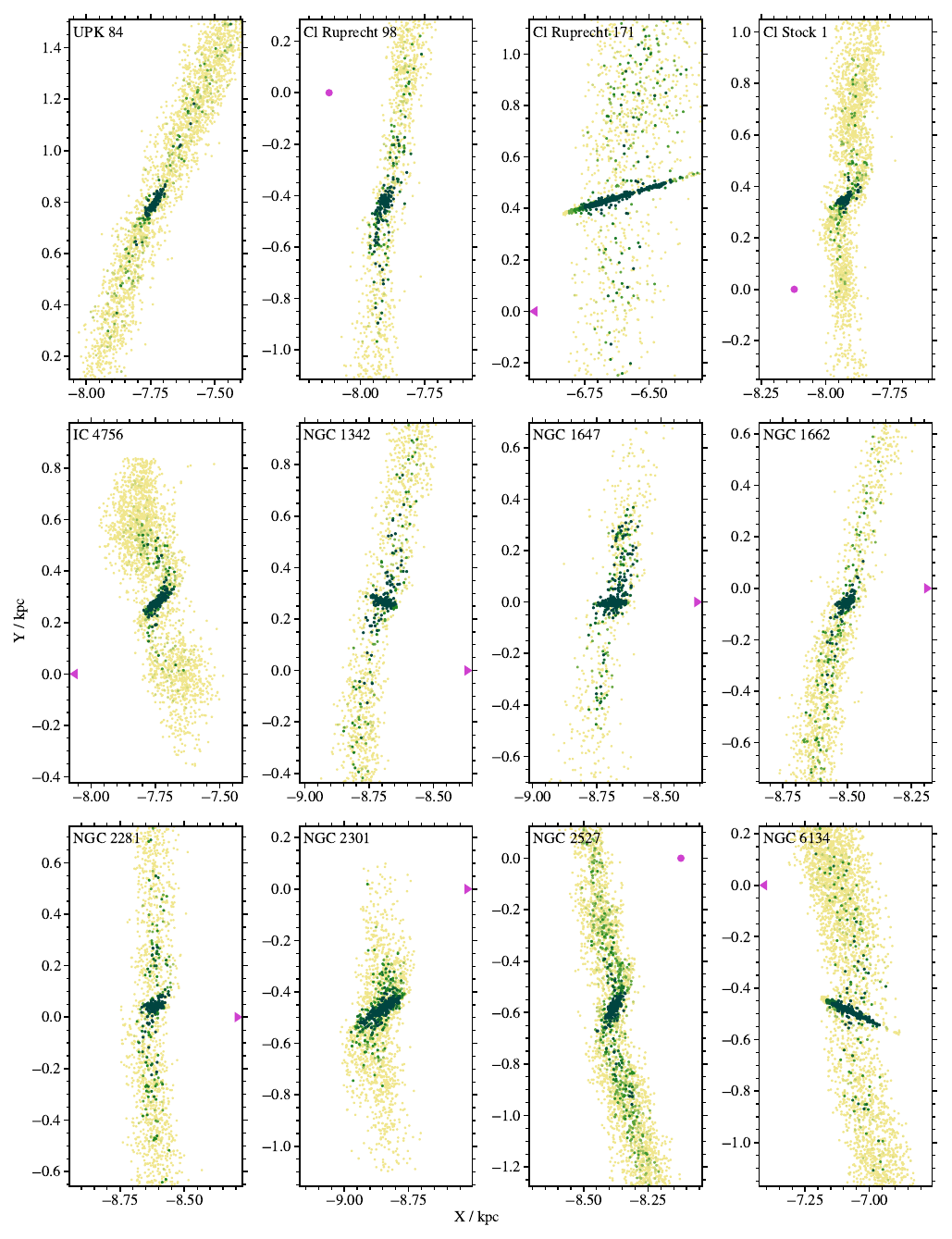}
    \caption{A selection of 12 clusters from this work displayed in the Galactic XY plane. Darker coloured points show stars with higher membership probability. The purple markings show the position of the Sun. The Galactic centre is toward the right and the Galaxy rotates towards positive $Y$ values. Stretching of the cluster core, most notable in Ruprecht 171 and NGC 6134, is due to distance uncertainties of individual stars which smear the distribution along the line-of-sight.}
    \label{fig:selection}
\end{figure*}

We provide the analysis of $476$ open clusters and a catalogue of $1\,026\,079$ stars found in their tidal tails. A small selection of clusters showing the diversity of found tidal tails is plotted in Figure \ref{fig:selection}

We also produce an array of diagnostic plots for each cluster (like Figures \ref{fig:hyades}, \ref{fig:ngc2516}, \ref{fig:coin}, and \ref{fig:ngc752}). All diagnostic plots show the same panels described here: Black contours show the log-density obtained from the simulation of cluster dissolution. Points are the stars with the highest likelihood, whereas the colour marks the membership probability. The purple dot (where visible) is the position of the Sun. The top-left plot shows the members in the $XY$ plane. Below it is the view in the Galactic plane, where $D$ is the dimension along the major axis of the cluster (defined for each cluster in Table \ref{tab:clusters}). The histogram on the bottom of this panel shows star counts in $25\ \mathrm{pc}$ bins along the $L$ axis. Bins of different membership probabilities are stacked on top of each other, and the solid line is the star count obtained as the sum of all membership probabilities (e.g. two stars with a membership probability of $0.5$ count as a value of one on the black histogram). Gray-shaded rectangles above the histogram mark the region used to calculate the asymmetry of the cluster. Moving one panel down, the colour-magnitude diagram shows the isochrone as the solid line for the age given in the right of the panel (sourced from \citet{cantatgaudin20}) and for the Solar metallicity. The dashed line is its binary sequence. The gray background represents the CMD of the general area around the cluster. The black arrow on the right (where visible) is the mean reddening vector. Positions of stars on the CMD were corrected for reddening and extinction as given in \textit{Gaia} DR3. On the bottom-left is a panel showing the mass function of the cluster. The histogram is stacked in the same way as the one two panels above. Blue line is the fitted de Marchi mass function. It was fitted onto the black histogram in the region that is plotted as a solid line. Dashed blue line is the extrapolation of the fitted function. Red and orange symbols show the degree of segregation (see Section \ref{sec:mass_seg}) for stars of the same or higher mass than the position of the symbol. Colours denote two different selections of stars (mass segregation based on members with $P>0.9$ or with $P>0.66$). In the right column, we show the position of the stars on the sky. Grey silhouette marks the shape of the Milky Way. Below is the proper motion diagram. Second from the bottom is the plot of radial velocities as a function of the distance of the star from the cluster centre projected on the sky. The bottom-right plot shows the two projected velocities (defined in Section \ref{sec:space}).

\subsection{Radial velocities of cluster centres}
\label{sec:newrv}

When comparing simulated clusters with found cluster members we noted many clusters that have wrong radial velocities reported in the literature. We obtained the radial velocities of clusters from Simbad (CDS) which collects them from many different sources (see Section \ref{sec:cluster_params}). Errors are mostly caused by using a too small sample of stars and by low-precision spectroscopy with large radial velocity uncertainties. Because we do not use radial velocities to find cluster members, we are somewhat immune to the erroneous radial velocity assumed for the cluster. This holds at least for the cluster cores, while the shape and position of the tidal tail depend on the radial velocity as well. 

In addition to the literature values, we calculated our own radial velocities of all clusters with at least 10 members with membership probability $P>0.5$ within $5^\circ$ of the literature's cluster centre. When our radial velocity was more than $10\ \mathrm{km\,s^{-1}}$ and three standard deviations different from the literature value, we simulated the cluster dissolution again, now using our radial velocities. This was done for 43 clusters. Revised radial velocities are collected in Table \ref{tab:new_rv}. These clusters have no radial velocity source in the \textit{Clusters} table, and the value of \texttt{radial\_velocity} is the one we measured for the cluster in the initial analysis and then used for the dissolution simulation. It can differ from the value of \texttt{radial\_velocity\_measured}, which we calculate after the final analysis, with the repeated simulation. For every cluster with at least 10 members with existing \textit{Gaia}'s radial velocities with membership probability $P>0.5$ within $5^\circ$ of the literature's cluster centre, we combine \textit{Gaia's} radial velocity measurements. When combining the measurements, we reject one third of the most extreme values (to mitigate the problem of binaries) and then report the average weighted by the \textit{Gaia}'s uncertainty and the standard deviation of the remaining sample. These values are reported in the \textit{Clusters} table in columns \texttt{radial\_velocity\_measured} and \texttt{radial\_velocity\_std}, respectively.

\begin{table}
\caption{Clusters with revised radial velocities. Columns are: the name of the cluster, number of members with membership probability $P>0.5$ used to calculate the radial velocity, radial velocity given in Simbad, source of Simbad's radial velocity, and the radial velocity measured by us with one standard deviation uncertainty.}
    \centering
    \small
    \setlength{\tabcolsep}{2pt} 
    \begin{tabular}{lccp{2.8cm}c}
    \hline
    Cluster & $N$ & $v_\mathrm{r}$ (Simbad) & $v_\mathrm{r}$ source (Simbad) & $v_\mathrm{r}$ (this work)\\\hline
     & & $\mathrm{km\, s^{-1}}$ & & $\mathrm{km\, s^{-1}}$\\\hline
NGC 129 & 11 & -35.67 & \citet{carrera19} & -48.7 $\pm$ 5.8 \\
COIN-Gaia 1 & 84 & 2.69 & \citet{monteiro19} & -13.3 $\pm$ 2.53 \\
COIN-Gaia 2 & 31 & 2.32 & \citet{monteiro19} & -32.11 $\pm$ 2.66 \\
UBC 194 & 16 & -4.95 & \citet{dias21} & -27.6 $\pm$ 2.18 \\
NGC 1027 & 38 & -30.86 & \citet{dias21} & -45.36 $\pm$ 4.3 \\
UPK 296 & 25 & -12.66 & \citet{dias21} & -3.29 $\pm$ 2.55 \\
NGC 1582 & 68 & 18.0 & \citet{tarricq21} & -23.61 $\pm$ 4.81 \\
NGC 1708 & 37 & -25.77 & \citet{dias21} & -8.22 $\pm$ 3.94 \\
COIN-Gaia 12 & 30 & 2.0 & \citet{monteiro19} & -8.96 $\pm$ 3.05 \\
COIN-Gaia 13 & 143 & 2.37 & \citet{monteiro19} & -12.58 $\pm$ 1.85 \\
Czernik 25 & 9 & 33.53 & \citet{tarricq21} & 44.21 $\pm$ 0.59 \\
Ferrero 11 & 52 & 32.24 & \citet{dias21} & 19.96 $\pm$ 3.23 \\
UBC 73 & 7 & 32.86 & \citet{tarricq21} & 25.03 $\pm$ 5.75 \\
UBC 212 & 13 & 41.69 & \citet{tarricq21} & 60.94 $\pm$ 7.1 \\
NGC 2335 & 12 & 8.7 & \citet{conrad17} & 23.91 $\pm$ 7.87 \\
NGC 2374 & 59 & 27.2 & \citet{conrad17} & 48.32 $\pm$ 3.98 \\
NGC 2548 & 259 & 41.5 & \citet{conrad17} & 8.14 $\pm$ 1.5 \\
UPK 537 & 81 & -3.86 & \citet{dias21} & 18.25 $\pm$ 3.62 \\
Pismis 4 & 136 & 27.2 & \citet{loktin17} & 8.69 $\pm$ 2.68 \\
UPK 528 & 53 & 3.62 & \citet{tarricq21} & 25.19 $\pm$ 4.43 \\
Gulliver 57 & 9 & 5.42 & \citet{dias21} & 20.22 $\pm$ 2.95 \\
LP 2238 & 30 & 19.96 & \citet{tarricq21} & -3.24 $\pm$ 3.98 \\
LP 1540 & 27 & 24.98 & \citet{dias21} & -22.48 $\pm$ 1.22 \\
ESO 130 08 & 47 & -38.52 & \citet{conrad14} & -13.86 $\pm$ 6.16 \\
Gulliver 58 & 39 & 9.86 & \citet{dias21} & -3.36 $\pm$ 7.48 \\
Loden 1194 & 42 & 2.0 & \citet{conrad17} & -27.6 $\pm$ 3.02 \\
NGC 5925 & 41 & -37.37 & \citet{dias21} & -21.01 $\pm$ 3.61 \\
NGC 6025 & 216 & -7.66 & \citet{gaia18} & 7.18 $\pm$ 3.84 \\
ASCC 88 & 33 & 1.8 & \citet{conrad17} & -16.18 $\pm$ 2.93 \\
UBC 95 & 23 & 2.26 & \citet{tarricq21} & -14.0 $\pm$ 9.11 \\
NGC 6494 & 403 & -31.73 & \citet{carrera19} & -9.41 $\pm$ 4.03 \\
UPK 5 & 55 & 7.81 & \citet{dias21} & -6.03 $\pm$ 3.18 \\
Ferrero 1 & 56 & 15.51 & \citet{tarricq21} & -5.06 $\pm$ 3.86 \\
NGC 6664 & 18 & -4.44 & \citet{dias19} & 16.01 $\pm$ 6.21 \\
UPK 13 & 56 & 24.65 & \citet{tarricq21} & 6.92 $\pm$ 6.25 \\
Czernik 38 & 19 & 57.39 & \citet{tarricq21} & 24.64 $\pm$ 8.11 \\
Ruprecht 145 & 150 & -8.4 & \citet{conrad17} & -54.72 $\pm$ 1.89 \\
UPK 54 & 79 & 47.07 & \citet{tarricq21} & -4.06 $\pm$ 4.05 \\
UPK 12 & 64 & -40.75 & \citet{dias21} & 3.4 $\pm$ 3.53 \\
UPK 45 & 108 & 26.6 & \citet{tarricq21} & -2.14 $\pm$ 6.01 \\
Alessi 44 & 163 & -62.2 & \citet{conrad17} & -9.65 $\pm$ 3.43 \\
NGC 7209 & 138 & -9.7 & \citet{conrad17} & -20.52 $\pm$ 3.27 \\
Alessi 37 & 167 & -23.35 & \citet{carrera19} & -13.08 $\pm$ 2.3 \\

         \hline
    \end{tabular}
    
    \label{tab:new_rv}
\end{table}

\subsection{Mass Functions}

To fit a mass function for hundreds of clusters one needs to parameterise it with a flexible function, yet simple and practical to fit. We chose the de Marchi mass function \citep{demarchi05, demarchi10}, a tapered power law of the form
\begin{equation}
    \frac{dN}{dm}\propto m^\alpha \left[ 1 - \exp\left(-\left(\frac{m}{m_\mathrm{c}}\right)^\beta\right) \right].
\end{equation}
The mass function is defined by the index of the power law ($\alpha$), describing the Salpeter-like mass function for most massive stars, characteristic mass $m_\mathrm{c}$, and the tapering exponent $\beta$ that describes the mass function at the low mass end, below $m_\mathrm{c}$.

We fit the observed mass distribution of every cluster with the de Marchi mass function with some additional restrictions: the integrated mass function must be finite, the index of power law must ($\alpha$) be close to $-2.4$, and the characteristic mass ($m_\mathrm{c}$) must be between $0.25$ and $0.85\ M_\odot$. To fit the mass function we use stars with all membership probabilities. The beauty of properly normalised membership probabilities is that we can add them together to get the star counts weighted with the membership probability. For example, ten stars with a membership probability of 0.1 together count for one star.  This means that we do not have to truncate the star counts at some membership probability. The benefit is particularly obvious in mass functions. The mass function for Hyades or NGC 752 in Figures \ref{fig:hyades} and \ref{fig:ngc752} is completely unrealistic if only high membership probability stars are used. When we add star counts for all stars, including low probability members, the mass functions look as expected (clear peak at around $0.7\ M_\odot$ and power law index of around $-2.4$). 

\subsection{Morphology parameters}

We calculated or fitted some parameters that describe the shape of the cluster and its tidal tails. These morphology parameters do not necessarily correspond to any physical parameters and must be used with caution when interpreted as such. We mainly use them to find clusters with particular properties in the database.

\subsubsection{Tail lengths}

Tidal tails stretch along the orbit of the cluster, which at scales of one kpc can hardly be approximated with a simple curve. We avoid using a different complicated orbit in the analysis of the obtained distribution of stars in each cluster. Instead we define a straight line called a \textit{major axis} that traverses the centre of the cluster and the extreme ends of both tidal tails. The major axis is calculated by fitting a line through the positions of cluster members in the $(X, Y, Z)$ space. Stars are weighted by their membership probabilities, and only stars in the top 10 percentile of the distance from the cluster centre are considered. This asures us that the major axis is fitted through the tails' ends. The parameters of the major axis are given in the \textit{Clusters} table. We use the major axis when it is beneficial to project the position of stars onto a single line, like in the calculation of the length of the tidal tails or cusp index in the next section, or when plots with the same orientation of the tails must be made for several clusters.

We calculate the total length of the tidal tails as the length along the major axis between the most extreme points with at least one star per $25\ \mathrm{pc}$ in star counts calculated as the sum of all membership probabilities projected onto the major axis, i.e. the distance between the extreme bins with the value of $\geq 1$ on the black histogram in the second panel on the diagnostic plots. Note that the tidal tails are curved, so the length of the tails measured along the orbit would be slightly larger. 

\subsubsection{Cusp index}

The cusp index parameterises the shape of the cluster system close to the core. Small numbers ($\sim 0.3$) indicate a concentrated and distinct core with a steep density drop toward the tidal tails (e. g. Hyades in Figure \ref{fig:hyades}). Large numbers ($\sim 1.5$) indicate a bulky cluster with a gradual density drop toward (usually short) tails (e.g. COIN-Gaia 13 in Figure \ref{fig:coin}). 

Cusp index (\verb+cusp_index+ in the \textit{Clusters} table) is calculated by fitting a generalised normal distribution to the logarithm of star counts (obtained as the sum of all membership probabilities) along the system's major axis. The generalised normal distribution has a form
\begin{equation}
\label{eq:general_norm}
    N(x)=A \frac{\beta}{2\alpha\Gamma(1/\beta)} \exp\left( -\left(\frac{|x-\mu|}{\alpha}\right)^\beta \right),
\end{equation}
where $\Gamma$ is the gamma function. Mean value $\mu$ is always set to zero, and $A$, $\alpha$, and $\beta$ are free parameters, the latter being the cusp index. 

Because the shape of the system can be distorted by the combination of distance uncertainties and projection to the major axis, we only calculate the cusp index for clusters closer than $1\ \mathrm{kpc}$.

\subsubsection{Tail asymmetry}

Tail asymmetry is calculated as the ratio of the number of stars in tidal tails on each side of the cluster centre, more precisely as:
\begin{equation}
    A=\frac{N(\alpha'<d_{\mathrm{m}}<500\ \mathrm{pc})}{N(-500\ \mathrm{pc}<d_\mathrm{m}<-\alpha')},
\end{equation}
where $\alpha'=0.5\,\alpha^2$, and $\alpha$ is defined in Equation \ref{eq:general_norm} and $d_\mathrm{m}$ is the position along the major axis. By counting stars from some distance away from the cluster centre we avoid making the asymmetry index too dependent on the star counts in the cluster core. Cutting off the star counts at $500\ \mathrm{pc}$ from the cluster centre we avoid the inclusion of stars at the end of the tails, where a large number of stars with small and possibly unreliable membership probabilities can add up to a significant star count. For clusters without the fitted $\alpha$ we assume a value of $\alpha=100\ \mathrm{pc}$.

\begin{figure}
    \centering
    \includegraphics[width=0.95\columnwidth]{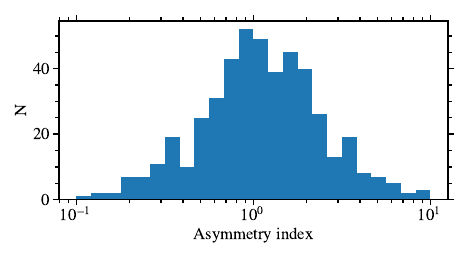}
    \caption{Distribution of asymmetry indexes of all clusters analysed in this work. }
    \label{fig:asym}
\end{figure}

Note that the tails that pass close to the Sun might be poorly surveyed by us and can therefore show some asymmetry due to incompleteness. 

We find that the distribution of asymmetries of tidal tails is normally distributed around the index $A=1$, so neither leading nor trailing tails are unevenly populated by stars. The distribution of asymmetry indexes is illustrated in Figure \ref{fig:asym}. One third of clusters have an asymmetry index such that one tail contains more than twice the number of stars as the other tail.

\subsubsection{Mass segregation index}
\label{sec:mass_seg}

We calculate the degree of mass segregation from the minimum spanning tree method as described in \citet{allison09}. The size of a minimum spanning tree is calculated for stars with mass larger than some value (mass ceiling). The size of this tree is compared with the average size of a minimum spanning tree calculated from the same number of randomly selected stars. If the two sizes are different, this indicates that the distribution of massive stars is different from the mean distribution of stars. 

We calculate the degree of mass segregation for a range of mass ceilings. The results are plotted on the bottom left panel in the diagnostics plots. The degree of mass segregation can be calculated for stars with different membership probability thresholds. We chose thresholds of $P>0.9$ and $P>0.66$ for the plots. Note that membership probabilities of individual stars cannot be used in the method described in \citet{allison09}, so we chose two thresholds that might represent the majority of stars with high membership probability. If there is mass segregation in the cluster system, the degree of mass segregation is expected to increase as a function of the mass ceiling. In the \textit{Clusters} table we indicate a possible trend with a single value: the ratio of the highest to the lowest degree of mass segregation.

We only report the mass segregation index when mass segregation is calculated with $P>0.9$ stars. If there are no high-probability stars in the tails, the mass segregation index can be recalculated using any probability threshold from the masses of stars reported in the \textit{Members} table. It is not possible to use membership probabilities to combine the contribution of many low-probability stars in this case, because the minimum spanning tree can only be constructed for distinct stars. 

\section{Discussions}
\label{sec:discussion}

\subsection{Clusters previously analysed in the literature}
\label{sec:literature}

\subsubsection{Hyades}
\label{sec:hyades}

Hyades have some of the longest known tidal tails of any open cluster. Tails have been found with several different methods \citep{meingast19, roser19, jerabkova21}. Hyades is one of the most convenient clusters for finding tidal tails because they are old enough to have developed long tails and they are one of the nearest clusters, so distances are precise, and radial velocities are available for many stars. However, no work in the literature attempts to estimate membership probabilities for Hyades stars, only finding members with the maximum likelihood.

\begin{figure}
    \centering
    \includegraphics[width=0.99\columnwidth]{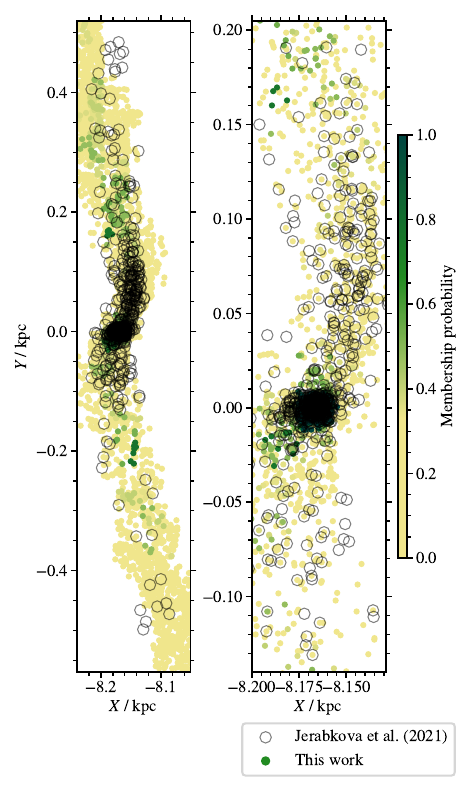}
    \caption{Comparison between stars found in this work (solid symbols) and by \citet{jerabkova21} (black circles) in Hyades and its tidal tails. Only stars with $G<17.5$ in \citet{jerabkova21} are shown.}
    \label{fig:hyades_xmatch}
\end{figure}

Our results for Hyades in general agree with the literature; we are able to find the same stars as in the literature, although the number of stars can differ depending on the likelihood threshold we adopt. Our goal is not to find the best likelihood threshold but to assign membership probabilities to stars. In Figure \ref{fig:hyades_xmatch} we show a detailed comparison of our results with \citet{jerabkova21}. Unfortunately, the stars outside the cluster core that we have in common with \citet{jerabkova21} are mostly low-probability members. The stars just outside of the cluster core are the ones where probability is the hardest to estimate, as they cover a large region in the parameter space that we then bin into small segments. In such regions even a small variation in the \textsc{Galaxia} star-counts can change the membership probability by a lot. Hyades is the nearest open cluster to the Sun, so this effect is the largest. Consequently, the Hyades is not the best cluster to compare the results of our method with the ones using maximum likelihood to define cluster members. 

Stars with membership probability $P>0.5$ found by us stretch for $800\ \mathrm{pc}$ between the tips of the tidal tails, the same length as uncovered in \citet{jerabkova21}. We also provide a selection of less probable members up to $750\ \mathrm{pc}$ from the core in each tail. The probability of finding a member at this extreme distance is around 5 stars per $25\ \mathrm{pc}$ of the tail length. This probability rises to around 10 stars per $25\ \mathrm{pc}$ at $500\ \mathrm{pc}$ distance from the core. The number of stars with nonzero likelihood at these distances is an order of magnitude higher than the number of expected member stars, so finding reliable members in these regions is hard and would require careful use of radial velocities and ideally chemical information as well. 

We also observe asymmetry in star counts close to the cluster core, same as \citet{jerabkova21}. However, the excess of stars in the leading tail is only apparent if membership probabilities are ignored. If counting only high-probability members, or summing all the membership probabilities, the asymmetry is reversed with more stars in the trailing tail. 

Hyades is one of the hardest clusters for determining membership probabilities. Their stars occupy a huge parameter space, so statistics is completely governed by low number star counts. When calculating likelihoods we can mitigate this by using more stars in the cluster dissolution simulation. However in the Galaxia simulation, this is not possible, as we already oversample the simulated population by a factor of five, and any significant increase in oversampling would need too much computer memory. The \textsc{Galaxia} simulation also lacks any fine structure that might affect star-count differences between observation and simulation close to the Sun. The same problem occurs with a handful of other clusters that are very close to the Sun or have tidal tails that pass close to the Sun. In the latter case, only the region closest to the Sun might be affected. 

\subsubsection{NGC 2516}

\begin{figure}
    \centering
    \includegraphics[width=0.95\columnwidth]{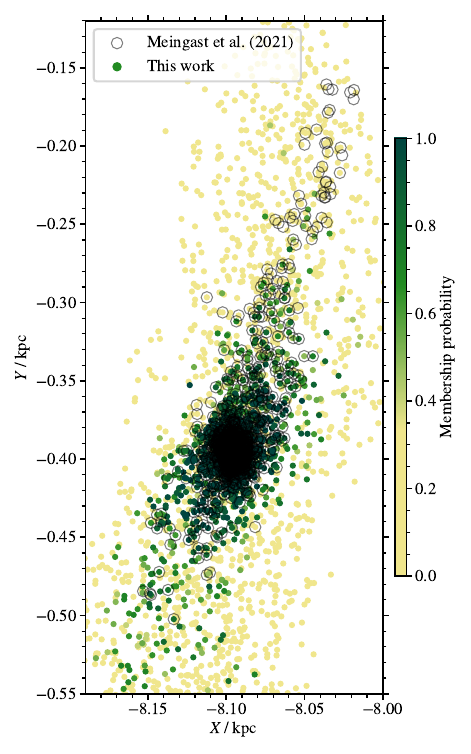}
    \caption{Comparison between stars found in this work (solid symbols) and by \citet{meingast21} (black circles) in NGC 2516 and its tidal tails. Only stars with $G<17.5$ in \citet{meingast21} are shown. There are more high probability members found by us outside of the plotted range (see Figure \ref{fig:ngc2516}).}
    \label{fig:ngc2516_xmatch}
\end{figure}

NGC 2516 is the cluster with the longest tidal tail in \citet{meingast21} measuring $380\ \mathrm{pc}$ from the end of the trailing to the end of the leading tidal tail. We find stars with membership probability $P>0.5$ stretching almost $500\ \mathrm{pc}$. From Figure \ref{fig:ngc2516_xmatch} we see that our members match well with stars found in \citet{meingast21} between $Y=-0.45$ and $-0.25\ \mathrm{kpc}$. Further in the leading tail the literature members only match with our low membership probability stars. The reason could be that closer to the Sun the tail covers a large region in the parameter space, so we cannot find any high membership probability stars there. However, in the trailing tail, we find high membership probability stars further away than \citet{meingast21}, consequently making our tidal tails more symmetric. Our complete results for NGC 2516 are illustrated in Figure \ref{fig:ngc2516}.

\subsubsection{COIN-Gaia 13}

\begin{figure*}
    \centering
    \includegraphics[width=0.95\textwidth]{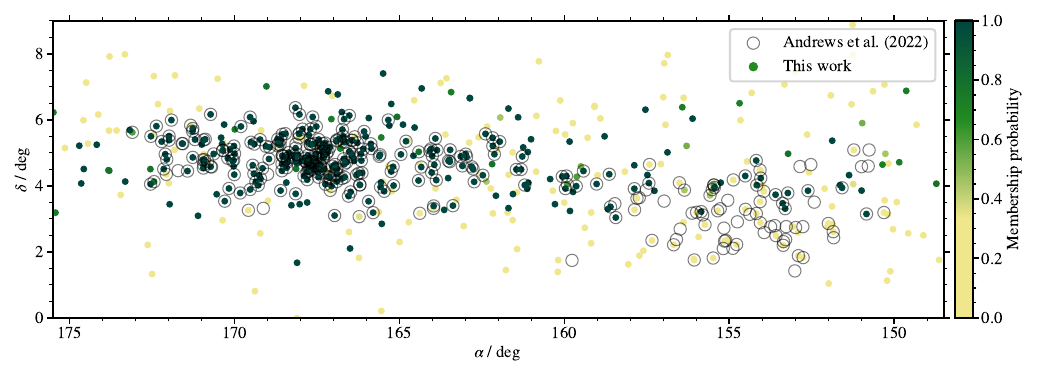}
    \caption{Comparison between stars found in this work (solid symbols) and by \citet{andrews22} (black circles) in COIN-Gaia 13 and its tidal tails. Only stars with $G<17.5$ in \citet{andrews22} are shown. There are more high probability members found by us outside of the plotted range (see Figure \ref{fig:coin}).}
    \label{fig:coin_xmatch}
\end{figure*}

COIN-Gaia 13, or Theia 456 is the cluster with the longest confirmed tidal tail in \citet{kounkel19} \citep{andrews22} spanning $200\ \mathrm{pc}$. The cluster is completely dissolved, showing no clear core, although a concentration of stars is observed at around $\alpha=83.186^\circ$, $\delta=42.087^\circ$, $\mu_\alpha \cos \delta=-3.828\ \mathrm{mas\, yr^{-1}}$, $\mu_\delta=-1.676\ \mathrm{mas\, yr^{-1}}$, and$\varpi=1.927\ \mathrm{mas}$ \citep{cantatgaudin20}. We detect stars with membership probability $P>0.5$ in a $700\ \mathrm{pc}$ long region stretching $50^\circ$ on the sky. Figure \ref{fig:coin_xmatch} shows that our member selections match well, although on the western part of the tail, the \citet{andrews22} stars match with a mix of our high and low membership probability stars. We also detect the same asymmetry as \citet{kounkel19} and \citet{andrews22}, showing more stars in the leading tail. 

Note that in COIN-Gaia 13 we found two giant stars with high membership probability. Based on their position in the HR diagram (Figure \ref{fig:coin}), they are not likely to be cluster members. They indeed have a negligible likelihood when their radial velocities are taken into account. Such occurrences are expected for stars with the membership probability lower than one. Our complete results for COIN-Gaia 13 are illustrated in Figure \ref{fig:coin}.

\subsubsection{NGC 752}

\begin{figure*}
    \centering
    \includegraphics[width=0.95\textwidth]{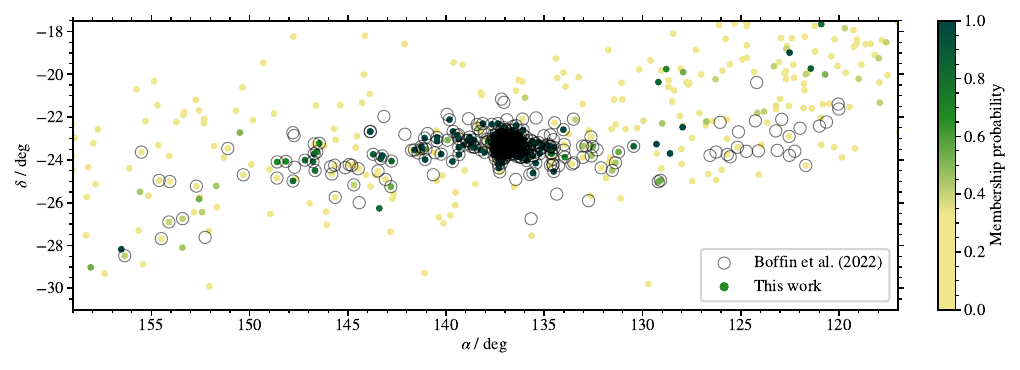}
    \caption{Comparison between stars found in this work (solid symbols) and by \citet{boffin22} (black circles) in NGC 752 and its tidal tails. Only stars with $G<17.5$ in \citet{boffin22} are shown. There are more high probability members found by us outside of the plotted range (see Figure \ref{fig:ngc752}).}
    \label{fig:ngc752_xmatch}
\end{figure*}

\citet{boffin22} claim that they possibly found stars in tidal tails of NGC 752 stretching along several kpc. It is unlikely that many stars several kpc from the cluster core are indeed members, as we cannot find any high membership probability stars at these distances. Hence in Figure \ref{fig:ngc752_xmatch} we only compare a smaller sample of stars found by \textsc{dbscan} in \citet{boffin22} with ours. Both our selections of stars match well close to the cluster core, but in the western tail, we cannot confirm most stars found by \citet{boffin22}. However, in the eastern tail, the match is very good and most high probability members found by us are also included in \citet{boffin22}. We also find stars with membership probability $P>0.5$ along a $1.5\ \mathrm{kpc}$ long stretch of the tidal tails. However, many of these stars that have radial velocities measured, would be excluded as high-likelihood members based on their radial velocities. In this case, we conclude that stars found by us at the extreme distances in the tidal tails of NGC 752 have overestimated membership probabilities. Our complete results for NGC 752 are illustrated in Figure \ref{fig:ngc752}.

\subsection{Completeness}

We analysed 476 clusters selected based on their age, distance, size, and availability of their mean radial velocity. From our input catalogue for clusters data \citep{cantatgaudin20} we removed 34 clusters due to no known radial velocity in the literature, after reducing the list of clusters by 223 clusters for being too small in terms of star counts. Therefore the clusters analysed in this work represent well the population of larger clusters in the solar neighbourhood, although not the complete population. Clusters with fewer members in \citet{cantatgaudin20} might be interesting for our analysis if the low star-counts are due to most stars already being moved into the tidal tails. 

In terms of the stellar content, our study is complete across the whole sky for all stars with $G<17.5$, as long as they are in the \textit{Gaia} DR3 catalogue. Particularly some bright stars might be missing, even if they are known cluster members. This should happen very rarely, so the effect for example on the mass function at the most massive end is negligible. 

We are confident that our results show observational evidence that almost all analysed clusters form tidal tails. The exceptions are most distant clusters where observational uncertainties blur the structures to such a degree that stars in the tidal tails cannot be found with any meaningful certainty. We refrain from giving a metric that would delineate clusters with recovered and non-recovered tidal tails. First reason is that there is no definition of what constitutes a tidal tail in terms of length or star density. Second reason is that the shape and properties of the tidal tails are determined by the stars found in them. Therefore we put our effort into computing the membership probabilities, which can be used for a variety of subsequent studies. It must be noted that a metric defining a well recovered tail would be extremely beneficial, as it would allow us to quantify which models and parameters constrain the position and shape of the tidal tails best, i.e. which models fit best the real distributions of stars.

\subsection{Dynamics of tidal tails}

Asymmetric tidal tails were first observed in Hyades \citep{meingast19, jerabkova21}, for which different causes were proposed. With our systematic study of 476 open clusters, we can confirm that asymmetric tidal tails are extremely common. However, we cannot confirm any bias in the asymmetry direction previously suggested in the literature \citep{kroupa22}. Leading and trailing tidal tails show the same degree of asymmetry (see Figure \ref{fig:asym}), even when we remove clusters where our survey is incomplete due to one of the tails passing close to the Sun.

Some clusters we analysed show evidence of epicyclic overdensities in their tails (best examples being Alessi 3 and IC 4651). For some clusters, more than one knot is observed. The structure of epicyclic overdensities depends on the orbit of the cluster, the position of the cluster on the orbit and the gravitational potential providing the tidal force \citep{just09}. Together with the position and velocity of the cluster the structure of overdensities offers a unique opportunity to study the gravitational potential of the Galaxy with some temporal resolution \citep{nibauer23}. Similar conclusions can be inferred from the position of tidal tails far away from the clusters. The latter would benefit from more model testing to find which gravitational potential in the cluster dissolution simulations produces the most reliable tidal tails at large distances. Some differences in the position of tidal tails can be seen in Figure \ref{fig:hyades_xmatch}, probably due to \citet{jerabkova21} and us using a different model of the Galactic gravitational potential (the main difference is the absence of the bar in \citet{jerabkova21}). The shape of the tidal tails might also depend on the mass of the cluster and the mass loss rate, which could be revealed by model testing. Model testing is out of the scope of this paper, as it would require more realistic n-body simulations. It would be best performed on a small number of clusters where we show that stars in the tidal tails can be found most reliably.

Epicyclic overdensities, tail orientations, and asymmetries are properties that arise due to many different physical processes. Before conclusions are made about what is responsible for the particular shape of a tidal tail it is best to verify the memberships through other means than just kinematics. Photometry and to some extent, radial velocities are readily available. However chemical analysis, like predicted Lithium abundance in younger clusters or some degree of chemical homogeneity among the found stars offers a good independent confirmation of cluster memberships \citep{casamiquela20, bouma21}.

\subsection{Using the catalogue}

\subsubsection{Table schema}
\label{sec:schema}
Resulting memberships and cluster parameters are collected in two tables (called \textit{Members} and \textit{Clusters}). The \textit{Clusters} table contains one entry per cluster with basic properties of the cluster, parameters used in our membership search and derived parameters. The \textit{members} table contains one entry per star per cluster with basic stellar parameters and our derived likelihoods and probabilities. 

\begin{table*}
    \centering
    \caption{Schema for the \textit{Clusters} table, listing all analysed clusters.}
    \begin{tabular}{llp{10.45cm}p{0.9cm}}
    \hline
         Column name & Units & Description & Data type\\
         \hline
        \verb+Cluster+ & -- & Cluster name as used in \citet{cantatgaudin20}. & str\\
        \verb+Cluster_Simbad+ & -- & Designation understandable by the Simbad (CDS) database. & str\\
        \verb+ra_icrs+ & $\mathrm{deg}$ & R.A. of cluster centre in decimal degrees in ICRS, as given in \citet{cantatgaudin20}. & float\\
         \verb+dec_icrs+ & $\mathrm{deg}$ & Declination of cluster centre in decimal degrees in ICRS, as given in \citet{cantatgaudin20}. & float\\
         \verb+X+ & $\mathrm{kpc}$ & X coordinate in the Galactocentric coordinate system. Sun is at $(8.122, 0.0, 0.0208)\ \mathrm{kpc}$. & float\\
         \verb+Y+ & $\mathrm{kpc}$ & X coordinate in the Galactocentric coordinate system. Sun is at $(8.122, 0.0, 0.0208)\ \mathrm{kpc}$. & float\\
         \verb+Z+ & $\mathrm{kpc}$ & X coordinate in the Galactocentric coordinate system. Sun is at $(8.122, 0.0, 0.0208)\ \mathrm{kpc}$. & float\\
         \verb+pmra+ & $\mathrm{mas}\,\mathrm{yr}^{-1}$ & Proper motion in R.A. of cluster centre multiplied by $\cos(\delta)$, as given in \citet{cantatgaudin20}. & float\\
         \verb+pmdec+ & $\mathrm{mas}\,\mathrm{yr}^{-1}$ & Proper motion in declination of cluster centre, as given in \citet{cantatgaudin20}. & float\\
         \verb+parallax+ & $\mathrm{mas}$ & Parallax of cluster centre as given in \citet{cantatgaudin20}. & float\\
         \verb+radial_velocity+ & $\mathrm{km\,s^{-1}}$ & Radial velocity of cluster centre from Simbad. & float\\
         \verb+radial_velocity_source+ & -- & Source of radial velocity. & str\\
         \verb+age+ & $\mathrm{Myr}$ & Age of the cluster from \citet{cantatgaudin20}. & float\\
         \verb+n+ & -- & Number of stars calculated as the sum of all membership probabilities. & float\\
         \verb+n_members+ & -- & Number of stars with reported membership probability. & int\\
         \verb+n_members_50+ & -- & Number of stars with membership probability $>0.5$. & int\\
         \verb+n_members_90+ & -- & Number of stars with membership probability $>0.9$. & int\\
         \verb+galaxia_com+ & -- & Mean completeness ($N_\mathrm{Gaia}/N_\mathrm{Galaxia}$) in the region of the cluster. & float\\
         \verb+A_G+ & mag & Mean extinction in G band for stars with membership probability $>0.5$ calculated using extinctions in \textit{Gaia} DR3. & float\\
         \verb+mass_func_index+ & -- & Index of the power law of the mass function ($\alpha$). & float\\
         \verb+mass_func_charm+ & $M_\odot$ & Characteristic mass of the mass function ($m_\mathrm{c}$). & float\\
         \verb+mass_func_taper+ & -- & Tapering exponent of the mass function ($\beta$). & float\\
         \verb+mass_segregation+ & -- & Mass segregation index calculated as $max(\Lambda)/min(\Lambda)$. Large values indicate that mass segregation was detected. & float\\
         \verb+cusp_index+ & -- & Index describing the shape of the cluster. Small value indicates a distinct core, large value indicates a gradual decrease in density from the core to the tails. & float\\
         \verb+tail_len+ & kpc & Length of tidal tail between most extreme points with $\geq10$ stars per $25\ \mathrm{pc}$. & float\\
         \verb+asym_index+ & -- & Asymmetry index calculated as the ratio of the number of stars on each side of the cluster core.  & float\\
         \verb+pa_d_axis+ & deg & Position angle of the major axis used to calculate projected positions \verb+d_axis+ in the \textit{Members} table& float\\
         \verb+radial_velocity_measured+ & $\mathrm{km\, s^{-1}}$ & Radial velocity of the cluster core measured by us. & float\\
         \verb+radial_velocity_std+ & $\mathrm{km\, s^{-1}}$ & Standard deviation of the radial velocities of all stars used to measure \verb+radial_velocity_measured+. & float\\

         \hline
    \end{tabular}
    \label{tab:clusters}
\end{table*}

\begin{table*}
    \centering
    \caption{Schema for the \textit{Members} table. 5D parameter space refers to $(x, y, z, \mu_{\parallel}, \mu_{\perp})$. 6D space adds radial velocity to these. }
    \begin{tabular}{llp{11.2cm}p{0.9cm}}
    \hline
         Column name & Units & Description & Data type\\
         \hline
         \verb+Cluster+ & -- & Cluster name as used in \citet{cantatgaudin20}. & str\\
         \verb+source_id+ & -- & \textit{Gaia} source identifier from \textit{Gaia} DR3. & long int\\
         \verb+ra_icrs+ & $\mathrm{deg}$ & R.A. in decimal degrees in ICRS, as given in \textit{Gaia} DR3. & float\\
         \verb+dec_icrs+ & $\mathrm{deg}$ & Declination in decimal degrees in ICRS, as given in \textit{Gaia} DR3. & float\\
         \verb+pmra+ & $\mathrm{mas}\,\mathrm{yr}^{-1}$ & Proper motion in R.A. multiplied by $\cos(\delta)$, as given in \textit{Gaia} DR3. & float\\
         \verb+pmdec+ & $\mathrm{mas}\,\mathrm{yr}^{-1}$ & Proper motion in declination, as given in \textit{Gaia} DR3. & float\\
         \verb+parallax+ & $\mathrm{mas}$ & Parallax as given in \textit{Gaia} DR3. & float\\
         \verb+radial_velocity+ & $\mathrm{km\,s^{-1}}$ & Radial velocity as given in \textit{Gaia} DR3. & float\\
         \verb+x+ & $\mathrm{kpc}$ & X coordinate in the Galactocentric coordinate system. Sun is at $(8.122, 0.0, 0.0208)\ \mathrm{kpc}$. & float\\
         \verb+y+ & $\mathrm{kpc}$ & Y coordinate in the Galactocentric coordinate system. Sun is at $(8.122, 0.0, 0.0208)\ \mathrm{kpc}$. & float\\
         \verb+z+ & $\mathrm{kpc}$ & Z coordinate in the Galactocentric coordinate system. Sun is at $(8.122, 0.0, 0.0208)\ \mathrm{kpc}$. & float\\
         \verb+d_axis+ & $\mathrm{kpc}$ & Position along the major axis of the cluster. The simulated cluster is used to calculate the orientation of the axis. & float\\
         \verb+l+ & $\mathrm{deg}$ & Galactic longitude. & float\\
         \verb+b+ & $\mathrm{deg}$ & Galactic latitude. & float\\
         \verb+d_2d+ & $\mathrm{deg}$ & Distance from the cluster centre projected onto the sky. & float\\
         \verb+proj_2d+ & $\mathrm{mas}\,\mathrm{yr}^{-1}$ & On-sky motion projected to the radial direction from the cluster (positive velocity means expansion). & float\\
         \verb+proj_perp_2d+ & $\mathrm{mas}\,\mathrm{yr}^{-1}$ & On-sky motion projected perpendicularly to the direction to the cluster centre. & float\\
         \verb+d_3d+ & $\mathrm{kpc}$ & Distance from the cluster centre. Parallax was directly inverted to obtain the distance from the Sun. & float\\
         \verb+proj_3d+ & $\mathrm{km\,s^{-1}}$ & Motion projected to the radial direction from the cluster (positive velocity means expansion). & float\\
         \verb+proj_perp_3d+ & $\mathrm{km\,s^{-1}}$ & Motion projected perpendicularly to the direction to the cluster centre. Can have any orientation. & float\\
         \verb+likelihood+ & -- & Probability that the simulation puts a star at this position in the 5D parameter space. Is not normalised. & float\\
         \verb+likelihood_6d+ & -- & Probability that the simulation puts a star at this position in the 6D parameter space (if \verb+radial_velocity+ exists). Is not normalised. & float\\
         \verb+probability_int+ & -- & Probability that the star is a cluster member calculated from the statistics for all stars within the same likelihood bin. Is normalised to $[0,1]$. & float\\
         \verb+probability_bin+ & -- & Probability that the star is a cluster member calculated from the statistics for a small region in 5D parameter space. Is normalised to $[0,1]$. & float\\
         \verb+phot_g_mean_mag+ & -- & G magnitude as given in \textit{Gaia} DR3. & float\\
         \verb+phot_bp_mean_mag+ & -- & BP magnitude as given in \textit{Gaia} DR3. & float\\
         \verb+phot_rp_mean_mag+ & -- & RP magnitude as given in \textit{Gaia} DR3. & float\\
         \verb+ebpminrp_gspphot+ & mag & Extinction in G band as given in \textit{Gaia} DR3. & float\\
         \verb+ebpminrp_gspphot+ & mag & Colour excess $E(BP-RP)$ as given in \textit{Gaia} DR3. & float\\
         \verb+ruwe+ & -- & RUWE value as given in \textit{Gaia} DR3. & float\\
         \verb+mass+ & $M_\odot$ & Mass of the star calculated from the isochrone for age given by \citet{cantatgaudin20} and Solar metallicity. Potential binaries are treated as single stars. & float\\
         
         \hline
    \end{tabular}
    \label{tab:members}
\end{table*}

\subsubsection{Using photometry for membership selection}

No photometric information apart from the cut in $G$ magnitude was used in the selection of possible cluster members or calculation of membership probabilities. We can then use HR diagrams to examine the performance of our methods, expecting the cluster members to be arranged on or close to the theoretical isochrone. The HR diagrams can be used to further clean the list of members for example to produce a better mass function at the high mass end, as a small number of stars misclassified as members can spoil the statistics (see for example Figure \ref{fig:ngc2516} where we found too many massive giants).

While we expect the found members to lie close to the theoretical isochrones, there are many caveats. Binary stars are found, which can lie up to 0.75 magnitudes above the isochrone. The ages of clusters can have high uncertainty and consequently, the correct isochrone cannot be determined well. Clusters between the age of 100 and $\sim300$ Myr are most affected because they don't have any pre-main sequence stars left, and at the same time only a small number of stars close to the turn-off point -- the position of both these stellar types being most sensitive to age. Finally, the biggest deviation from stars being nicely arranged on an isochrone is the differential interstellar extinction. Because we are finding stars that sometimes stretch along several kiloparsecs in space, the measured magnitudes are affected by a significantly varying range of extinctions. To mitigate this, we use \textit{Gaia}'s extinctions and reddenings for individual stars when plotting HR diagrams. An alternative would be to produce the PADOVA isochrones with a mean extinction applied, which would not solve the issue of differential reddening. Extinction and reddening are not always consistently measured across different regions on the sky or across different stellar types, which is most evident for clusters with high extinctions (for example cluster King 1). There is also an issue of theoretical PADOVA isochrones not matching well with observations (for example in Hyades between $2<BP-RP<3$). The precision of measured magnitudes is not an issue for stars with $G<17.5$, apart from the stars with bad quality data. 

\subsubsection{Caveats}
\begin{itemize}
    \item Do not use likelihood to select members, unless you will be using an additional algorithm for finding the most probable members like clustering algorithms, or use more parameters, like radial velocity, photometry, etc.
    \item The catalogue is not complete. While we try to provide a complete selection of members above some likelihood threshold, we fail to do so in cases where tidal tails pass close to the Sun. Those stars occupy a large region in the parameter space, which makes it difficult to calculate likelihoods or membership probabilities.
    \item Stars can have a large likelihood (or even probability) to be in more than one cluster. This happens when nearby clusters move along similar orbits. In practice, such stars can be assigned to the cluster with the highest likelihood. We leave stars with multiple cluster memberships in the tables for the user to decide what to do with them. There are almost $12\,000$ such stars, but almost always one cluster has a much higher membership probability (most typically by a factor of 10). For a few percent of stars where the membership probability is similar for bots clusters, both probabilities are low (almost always $<0.2$). Stars can also belong to clusters not analysed in this work (for example UPK 537 contains stars from (probably co-natal) UBC 480).
    \item We report membership probabilities computed by two different methods (see Section \ref{sec:intbin}). We recommend the integrated probability to be used over the binned probability, as the latter is more susceptible to small number statistics and can thus vary a lot in the most extreme parts of the tidal tails.
    \item Membership probabilities are always calculated for small groups of stars. While such a probability might be reliable for the whole ensemble of stars, one must be careful when interpreting the value for each individual star. E.g. a group of high probability stars (but with probability $<1$) probably includes members and some nonmembers. Concluding that all high probability stars are indeed members of a cluster is therefore wrong. See Section \ref{sec:robust} for more discussion.
    \item We include in our analysis all \textit{Gaia} stars, regardless of the quality of the measured parameters (given by RUWE, for example). This is necessary to achieve completeness which is necessary to reliably measure membership probabilities. User must be careful to avoid the stars with poor-quality data, as their parameters reported in this work might be erroneous. 
\end{itemize}

\section{Conclusions}
\label{sec:conclusions}

In the third data release of \textit{Gaia}, the astrometric parameters have become precise enough to enable the discovery of stars that have escaped the potential of the core of open clusters. These stars form tidal tails under the influence of the Galactic gravitational forces. Tidal tails of open clusters were long predicted to exist, but we offer the first systematic study that shows that they are common and can be recovered to a large extent. 

We have derived normalised membership probabilities, which give the probability that a star is a cluster member, as opposed to likelihoods -- usually used in the literature -- that give the probability that cluster members are found at some location. We show that correctly normalised membership probabilities are extremely beneficial for statistical studies of tidal structures in regions where cluster membership cannot be determined with absolute certainty. This aids us in recovering the extent of tidal tails and accounts for stars in tidal tails in mass functions for example. Rich and diverse structures that we observe in tidal tails of 476 analysed clusters indicate that many different physical processes can be studied in the future based on the structure of tidal tails. 

To perform a more reliable and precise search of tidal tails in future studies, it will be essential to use radial velocities, provided that the selection function for stars with available radial velocities can be derived. A better known selection function for fainter stars and for binaries would also be needed to extend the search toward fainter stars. To reliably estimate the membership probabilities we also seek the models of the Galactic stellar population that are more realistic on smaller scales. 

It should also be possible to include masses of stars into the likelihood calculation if some IMF and degree of mass segregation could be assumed. Any additional parameter that can increase the dimension of the parameter space is beneficial to the likelihood calculation. Stellar mass is already a parameter in the synthetic models of the stellar population, so calculating membership probabilities using stellar masses is not a problem. However, the IMF and mass segregation can depend a lot on the conditions during the star formation and on perturbations that a cluster encounters in its lifetime. Dedicated cluster-specific simulations would then need to be done instead of a one-fits-all type of simulation of a cluster dissolution that we used in this work. 

\begin{acknowledgements}
      JK is supported by the Slovenian Research Agency ARIS grants P1-0188. JK thanks Sanjib Sharma for all the useful discussions and for the use of the most recent version of the \textsc{Galaxia} code with \textit{Gaia}'s photometric filters. JK is most thankful to the anonymous referee, who provided many valuable comments and initiated several highly appreciated discussions that have undoubtedly enhanced this work. This work has made use of data from the European Space Agency (ESA) mission {\it Gaia} (\url{https://www.cosmos.esa.int/gaia}), processed by the {\it Gaia} Data Processing and Analysis Consortium (DPAC, \url{https://www.cosmos.esa.int/web/gaia/dpac/consortium}). Funding for the DPAC has been provided by national institutions, in particular the institutions participating in the {\it Gaia} Multilateral Agreement.
\end{acknowledgements}

\bibliographystyle{aa} 
\bibliography{bib} 

\begin{appendix} 

\section{Full parameter space plots}

\begin{figure*}
    \centering
    \includegraphics[width=\textwidth]{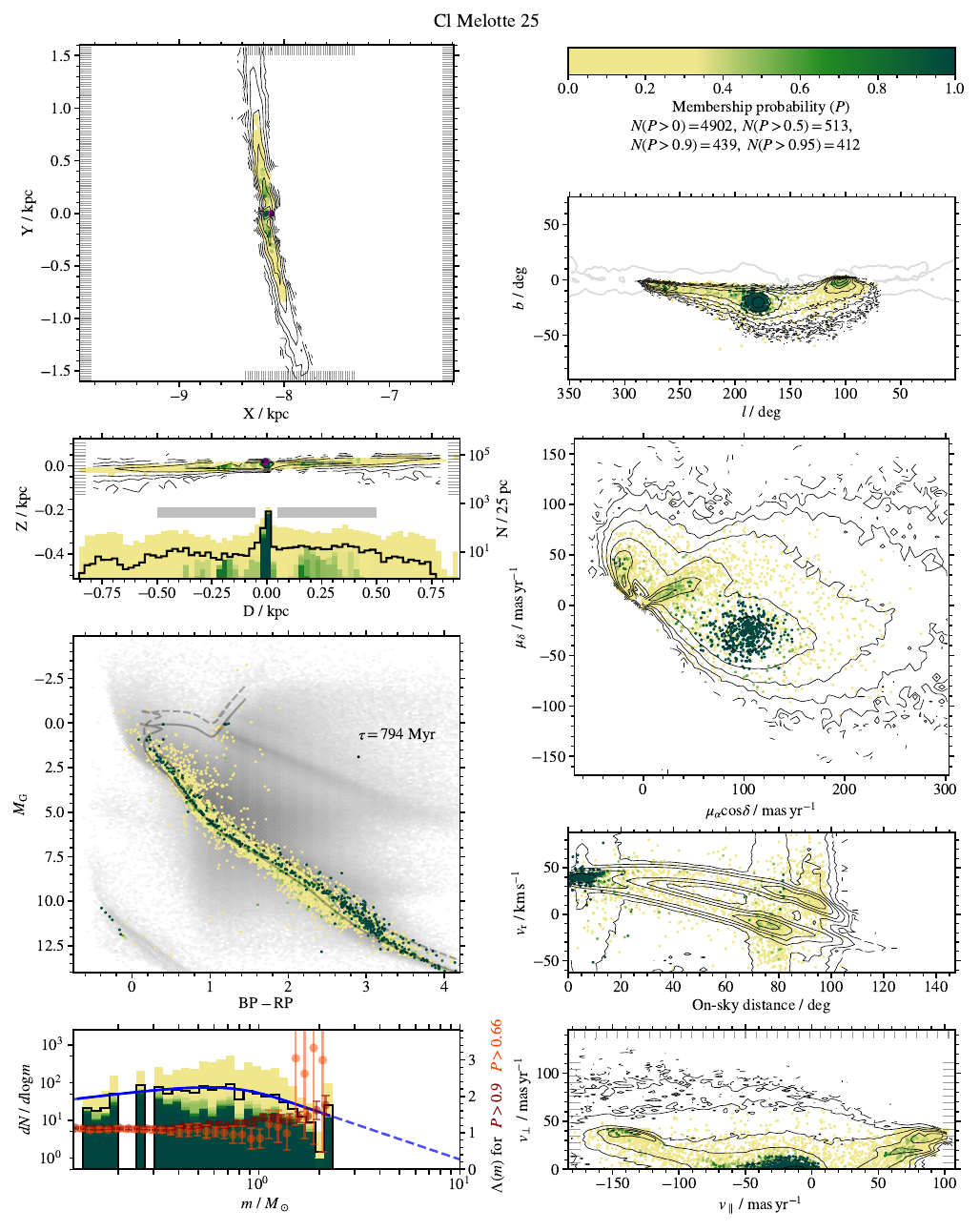}
    \caption{Membership probability of stars in Hyades displayed across the parameter space. See Section \ref{sec:results} for the description of this Figure.}
    \label{fig:hyades}
\end{figure*}

\begin{figure*}
    \centering
    \includegraphics[width=\textwidth]{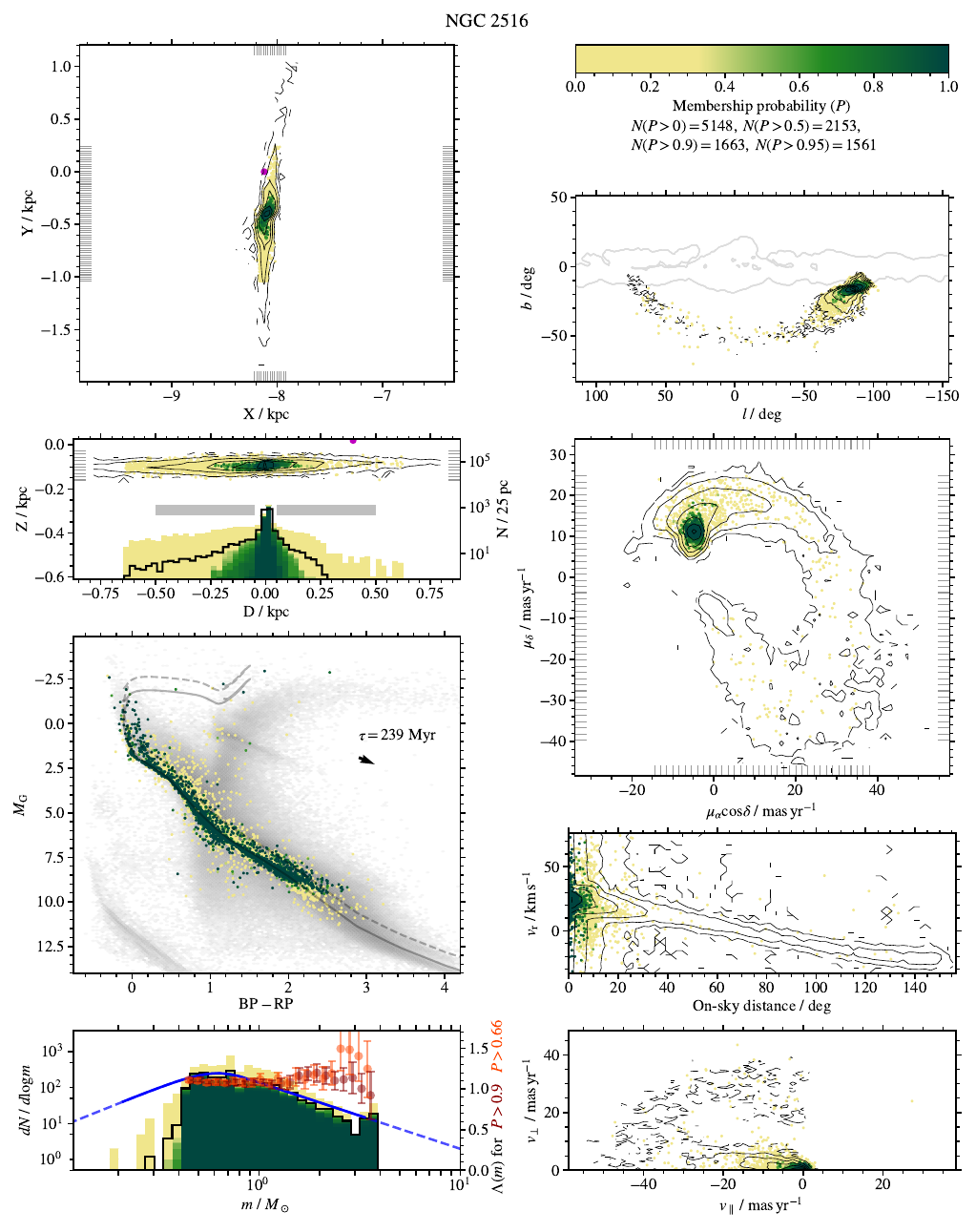}
    \caption{Membership probability of stars in NGC 2516 displayed across the parameter space. See Section \ref{sec:results} for the description of this Figure.}
    \label{fig:ngc2516}
\end{figure*}

\begin{figure*}
    \centering
    \includegraphics[width=\textwidth]{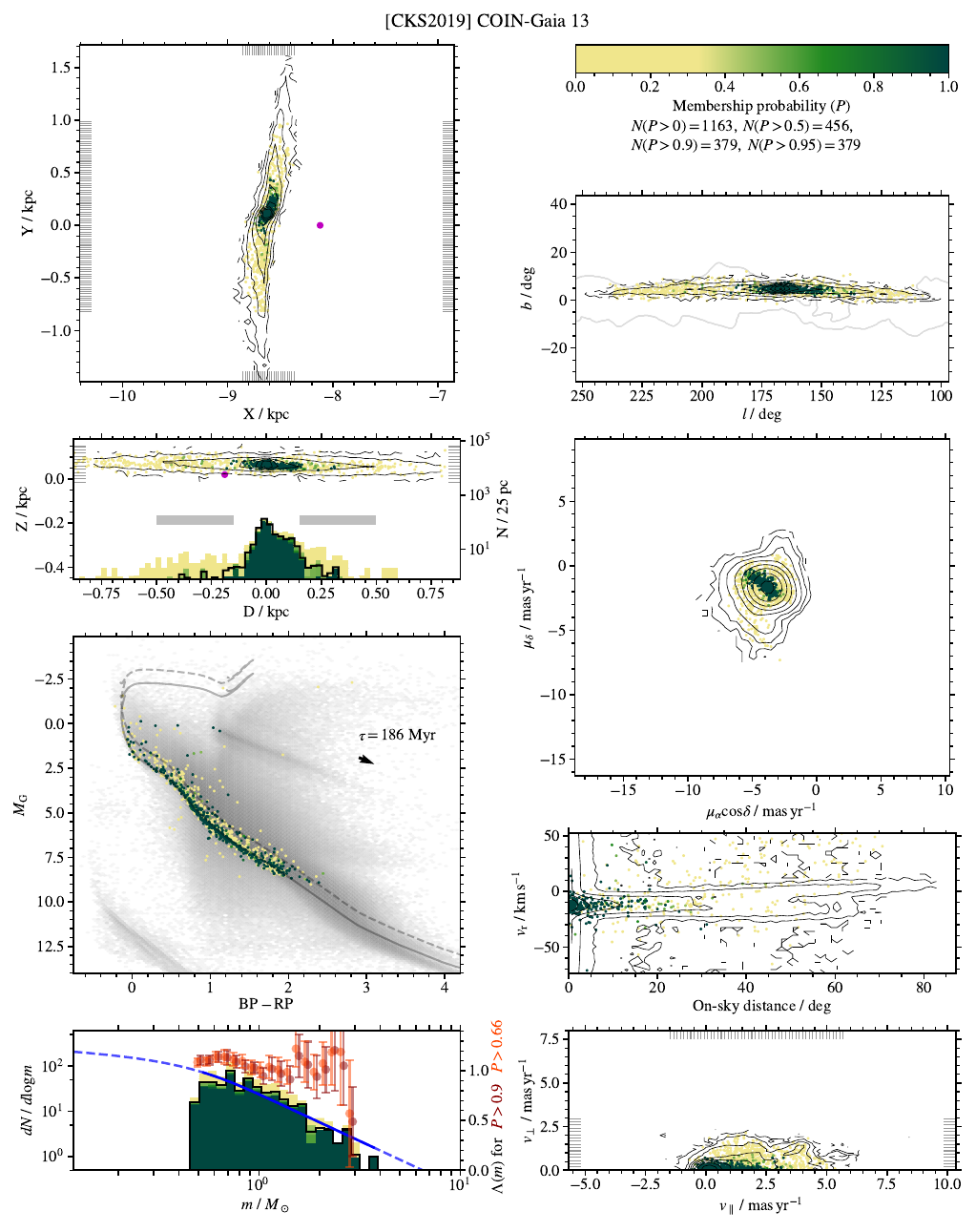}
    \caption{Membership probability of stars in COIN-Gaia 13 displayed across the parameter space. See Section \ref{sec:results} for the description of this Figure.}
    \label{fig:coin}
\end{figure*}

\begin{figure*}
    \centering
    \includegraphics[width=\textwidth]{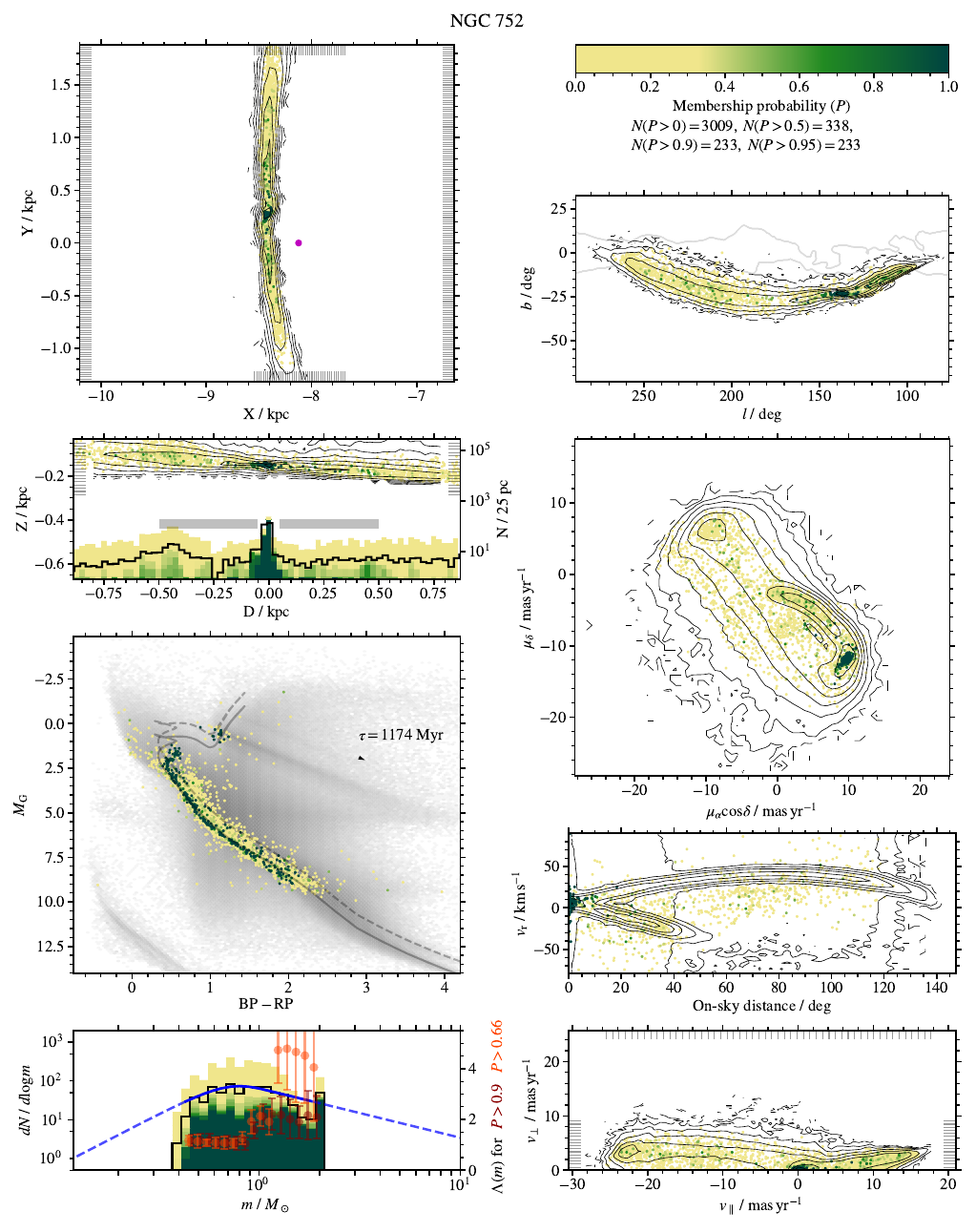}
    \caption{Membership probability of stars in NGC 752 displayed across the parameter space. See Section \ref{sec:results} for the description of this Figure.}
    \label{fig:ngc752}
\end{figure*}

Figures \ref{fig:hyades} to \ref{fig:ngc752} show the full parameter space plotted for four clusters discussed in more detail in Section \ref{sec:literature}.

\section{Accuracy of simplified cluster dissolution simulations}

\subsection{Two-body interactions}

Our simulations of cluster dissolution use an empirical gravitational potential that represents the mass distribution in an open cluster. Simulated stars are then massless particles that do not interact with each other and only feel the combined gravitational potential of the Galaxy and the cluster. This is a simplification that allows us to compute simulations faster and also imposes no selection functions (based on stellar masses or spectral types) that we would have to account for in the observed samples. In order to verify that a simplified simulation can reproduce the same structures that an n-body simulation would, we ran an independent n-body simulation of the cluster NGC 2516.

The n-body simulation initialises a cluster with a Plummer gravitational potential with a radius of $2\ \mathrm{pc}$, and 3500 particles with a Kroupa mass function (corresponding to a total mass of around $1300\ M_\odot$). The initial position of the cluster is such that the centre of the cluster ends at the observed position and velocity when simulated to the present time. The simulation did not contain any binary stars. We used the \textsc{AMUSE}\footnote{\url{https://www.amusecode.org/}} framework \citep{portegies09, portegies13, pelupessy13, portegies18} to combine the \textsc{BHTree} n-body simulation code with the same gravitational potential of the Galaxy as used in the simplified simulations. The simulation was repeated 50 times to produce enough samples to compare it to a larger simplified simulation.

\begin{figure}
    \centering
    \includegraphics[width=\columnwidth]{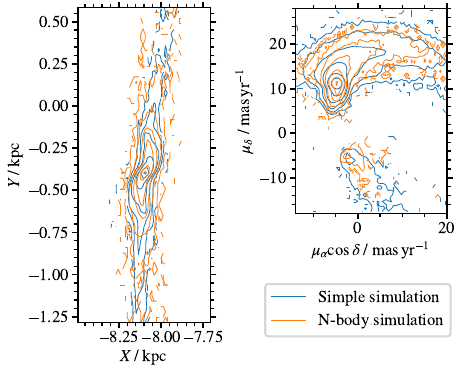}
    \caption{Comparison between the simulated tidal tails in a simplified simulation and in an n-body simulation. Simplified simulation (blue contours) is used throughout this work. A collisional n-body simulation (orange contours) of the same cluster NGC 2516 shows the same shape of tidal tails in the spatial (left panel) and kinematic (right panel) parameter spaces. The distributions plotted in this figure are not convolved with observational uncertainties. Contours are log-spaced.}
    \label{fig:nbody_comp}
\end{figure}

Comparison between a simplified simulation and an n-body simulation is shown in Figure \ref{fig:nbody_comp} for two spatial and two kinematic dimensions of cluster NGC 2516. Two simulations produce the exact same structures with the same positions and orientations in the parameter space. The only notable difference is a somewhat slower drop in density in the $XY$ plane for the n-body simulation. We conject that using such a simulation in the calculation of membership likelihoods would result in a similar final selection of stars, as low-likelihood stars do not make the selection as possible cluster members. Processes not simulated in either of the two simulations (like interactions with GMCs and spiral arms) probably have larger impact on the shape of the tidal tails than is the discrepancy between the two simulations displayed in Figure \ref{fig:nbody_comp}.

\subsection{Cluster mass}

Massless particles in our simplified simulations are contained in a Plummer potential that moves on a fixed orbit around the Galaxy. The mass of the Plummer potential, which represents the mass of the cluster decreases with time (see Section \ref{sec:cluster_params}).

\begin{figure}
    \centering
    \includegraphics[width=\columnwidth]{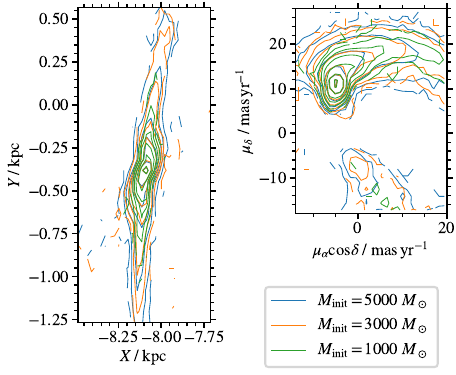}
    \caption{Comparison between simulations with different initial cluster masses. All simulations are made for cluster NGC 2516 with a cluster radius of $2\ \mathrm{pc}$. The distributions plotted in this figure are not convolved with observational uncertainties. Contours are log-spaced.}
    \label{fig:var_m}
\end{figure}

We tested whether the initial cluster mass has any effect on the shape of the tidal tails. Figure \ref{fig:var_m} shows three simulations for cluster NGC 2516 with different initial cluster masses. Initial masses of $M_\mathrm{init}=5000$ and $3000\ \mathrm{M_\odot}$ exhibit no significant difference, while the initial mass of $M_\mathrm{init}=1000\ \mathrm{M_\odot}$ produces slightly shorter tails.  Shorter tails are also reflected in the distribution of stars in the simulated proper motion plot (right-hand-side plots in figures in this section), where there are almost no stars around the convergent point at $(\mu_\alpha, \mu_\delta)=(0, -6)\ \mathrm{mas\, yr^{-1}}$.

\subsection{Cluster size}

\begin{figure}
    \centering
    \includegraphics[width=\columnwidth]{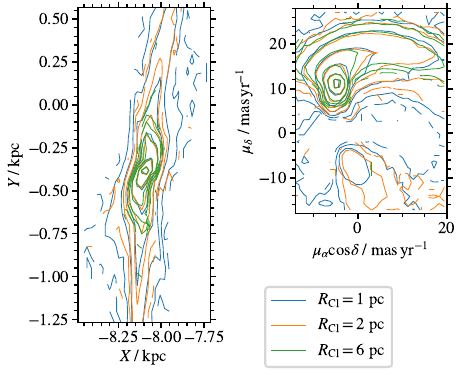}
    \caption{Comparison between simulations with different cluster radii. All simulations are made for cluster NGC 2516 with an initial mass of $5000\ \mathrm{M_\odot}$. The distributions plotted in this figure are not convolved with observational uncertainties. Contours are log-spaced.}
    \label{fig:var_r}
\end{figure}

While the mass of the Plummer potential decreases through the simulation, the radius of the Plummer potential is kept constant at $R_\mathrm{Cl}=2\ \mathrm{pc}$. Figure \ref{fig:var_r} shows the sensitivity of the simulation results on the varying cluster radius. The simulation of cluster dissolution was made with three cluster radii. Shown is the expected shape at the present time for cluster NGC 2516. The corona of the cluster is more puffed-up for smaller clusters, and the tails are slightly shorter in the case of $R_\mathrm{Cl}=6\ \mathrm{pc}$. The shape and orientation of the tidal tails are the same for all cases. We conclude that adopted cluster sizes and masses have negligible effect on the large-scale structure of the tidal tails unravelled in this work.

\section{Robustness of derived membership probabilities}
\label{sec:robust}

We derive membership probabilities under the assumption that the model for cluster dissolution is correct. The model assumes a Galactic gravitational potential, bar pattern speed, shape of the cluster and its initial velocity dispersion, \textit{Gaia}'s selection function, and a model for the Galactic stellar population -- all of which are findings of a broad part of modern astrophysics. We use models and parameters that are widely accepted as most probable values for state-of-the-art models. It is well beyond the scope of a single paper to verify the assumed models and explore their effects on the tidal tails of open clusters, let alone marginalise the probabilities calculated in this work over all possible variations of the assumed models. It is also impossible to include a number of secondary perturbations, like interaction of clusters with spiral arms and giant molecular clouds into the simulations, as their interaction with stars and clusters throughout the last 100s million years is largely unknown. While such perturbations have certainly affected some clusters and their tidal tails (signs of which we point out in this work), they would have to be studied case-by-case. Hence we emphasis that the results presented in this work are produced under the assumption of one set of models.

\begin{figure}
    \centering
    \includegraphics[width=\columnwidth]{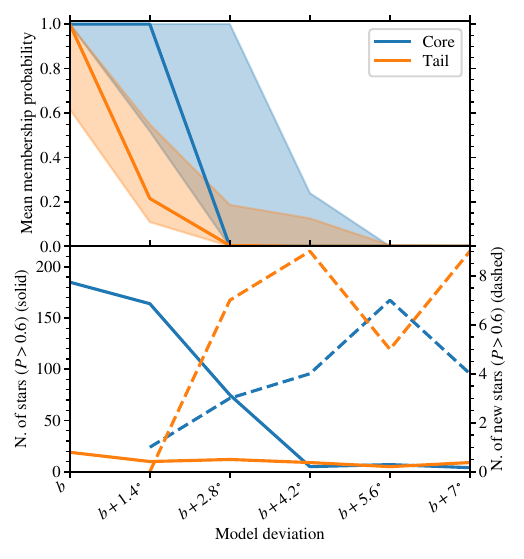}
    \caption{Star counts and membership probabilities in the core and tails of cluster Collinder 350 if simulated with different positions of the cluster. Top: Mean membership probability of stars that have membership probability $P>0.6$ in our original calculation after recomputing them with different cluster models. Shades show 16th and 84th percentiles and the silid lines show the 50th percentile. Bottom: Number of stars with $P>0.6$ in the core (blue) and tails (orange) is shown with solid lines for the original cluster model and five different deviations from the model. Dashed lines show the number of stars that do not appear in the selection made with the original cluster model. All models deviated from the original in the position of the cluster centre in galactic latitude.}
    \label{fig:div_pos}
\end{figure}

\begin{figure}
    \centering
    \includegraphics[width=\columnwidth]{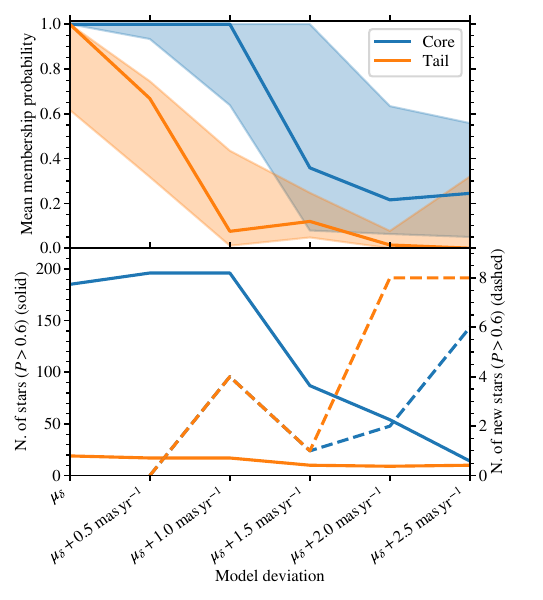}
    \caption{Same as Figure \ref{fig:div_pos}, but for model deviation in proper motion.}
    \label{fig:div_pm}
\end{figure}

In this section we explore the sensitivity of our selection of stars in tidal tails on the variations in underlying models. In this test we try to find tidal tails of imaginary clusters, i.e. clusters for which we made up some arbitrary positions and motions. We start with a real cluster and vary the proper motion and the position of the cluster centre in our simulation, until they represent a nonexistent cluster, well beyond the uncertainty of the parameters defining the cluster centre. Figures \ref{fig:div_pos} and \ref{fig:div_pm} quantify the results for one cluster (Collinder 350). We found members and recalculated their membership probabilities when pretending that the cluster centre is between $1.4^\circ$ and $7^\circ$ from the real one. First we traced how the membership probability of stars found in the original calculation (with cluster's centre unchanged) varies in subsequent computations with a shifted cluster centre. First, we selected all stars with membership probability above 0.6 in the original calculation. In the top panel of Figure \ref{fig:div_pos} we show the 16th to 84th percentile of recalculated memberships of these stars in subsequent simulations. Stars that initially had a high membership probability, become low membership probability stars or are completely absent in the computations with a shifted cluster centre. We also plot the total number of high probability stars ($P>0.6$ hereafter) in each individual computation (solid line, bottom panel of Figure \ref{fig:div_pos}) and a number of high probability stars that appear in computations with shifted cluster centre that do not exist in the original computation. These are expected to be stars that we serendipitously find to be high probability members, but are obviously not, as they only appear in computations with a shifted cluster centre. We repeated the analysis for the cluster centre at the original position, but this time with the proper motion of the cluster shifted between $0.5$ and $2.5\ \mathrm{mas\, yr^{-1}}$ as illustrated in Figure \ref{fig:div_pm}.

It is expected that by shifting the alleged cluster centre away from the real one, we loose the original cluster members. This transition happens at the shift of around $2^\circ$ or $1.5\ \mathrm{mas\,yr^{-1}}$ for the cluster core. This is around twice the standard deviation of the core's size in position and proper motion. The loss of original stars is more rapid in the tails, also because the membership probabilities in the tail are lower in average and closer to the threshold of $P=0.6$, as used in the aforementioned exercise. More informative are the star counts. The number of high probability stars in the tails drops from 20 in the original computation to 9 in the computations with the most extreme shift. Almost all stars in the tail at large shifts are stars with a zero membership probability in the original computation. Given that probabilities of stars found in the tails are usually less than one, we expect that some are just coincidental stars that we could not have rejected as cluster members using only five observables. By the same logic, it is expected that for a computation with a shifted cluster centre, we find stars that satisfy the condition to be a cluster member. The fact that the number of stars in the tidal tails drops by a significant fraction when we move the proposed cluster centre from the real one, confirms that the calculated probabilities are sensible. Although it must be noted, that the number of stars found in the tails of shifted alleged clusters is not zero. A drop from 20 to 9 stars indicates, that the stars in the tidal tails that had membership probabilities $>0.6$ for the computation for a real cluster should have somewhat lower probabilities. If 9 out of 20 stars are impostors, as the above experiment suggests, the membership probabilities of the initial 20 stars should be $0.55$ in average.

A test presented in this section would have to be repeated for a large number of clusters in order to get a good overview of the discussed behaviour elsewhere in the Galaxy. It must be noted that the number of stars we might find in imaginary clusters can depend on the position in the Galaxy and on the parameters of the cluster. A quick estimate that suggests that the probabilities might be slightly overestimated would likewise need a thorough analysis. However, it is expected that the probabilities derived from low number statistics are only our best estimates. Repeating the analysis for all clusters would be a computationally expensive exercise, which we are not able to perform.

\section{List of analysed clusters}

A list of all clusters is given in Table \ref{tab:all_clusters} with celestial coordinates and a number of found members. A table with all reported parameters is available online (see description in Section \ref{sec:schema} and Table \ref{tab:clusters}).

\begingroup

\begin{table*}
\caption{All analysed clusters. $N_{P>0.9}$ is the number of stars with membership probability $>90\%$. A table with more parameters is available online.}
    \label{tab:all_clusters}
\scriptsize
\setlength{\tabcolsep}{2pt} 
\renewcommand{\arraystretch}{0.39} 
\setlength{\columnsep}{1.0mm}
\begin{minipage}{\textwidth}
\begin{multicols*}{4}
\centering
    \begin{tabular}{p{1.8cm}ccc}
        \hline
        Cluster name & $\alpha\ /\ ^\circ$ & $\delta\ /\ ^\circ$ & $N_{P>0.9}$ \\
        \hline
    Blanco 1 & 0.853 & -29.958 & 411 \\
Gulliver 24 & 1.161 & 62.835 & 119 \\
UPK 237 & 5.352 & 65.495 & 124 \\
King 1 & 5.505 & 64.383 & 399 \\
UBC 184 & 5.76 & 59.451 & 70 \\
FSR 0496 & 6.656 & 64.01 & 310 \\
Stock 21 & 7.59 & 57.922 & 88 \\
NGC 129 & 7.607 & 60.213 & 0 \\
NGC 189 & 9.885 & 61.09 & 134 \\
NGC 188 & 11.798 & 85.244 & 560 \\
COIN-Gaia 1 & 11.933 & 66.769 & 176 \\
Alessi 1 & 13.343 & 49.536 & 61 \\
COIN-Gaia 2 & 15.06 & 55.409 & 116 \\
NGC 381 & 17.094 & 61.586 & 165 \\
NGC 433 & 18.798 & 60.133 & 127 \\
COIN-Gaia 30 & 21.08 & 70.574 & 134 \\
Collinder 463 & 27.031 & 71.738 & 531 \\
NGC 752 & 29.223 & 37.794 & 233 \\
NGC 743 & 29.614 & 60.13 & 151 \\
NGC 744 & 29.652 & 55.473 & 194 \\
UBC 44 & 31.11 & 54.359 & 42 \\
Stock 5 & 31.119 & 64.365 & 208 \\
Stock 2 & 33.856 & 59.522 & 1063 \\
UBC 419 & 34.694 & 58.924 & 65 \\
NGC 886 & 35.898 & 63.8 & 182 \\
UBC 194 & 37.686 & 48.305 & 108 \\
King 4 & 39.028 & 59.024 & 79 \\
Trumpler 2 & 39.232 & 55.905 & 289 \\
NGC 1039 & 40.531 & 42.722 & 498 \\
NGC 1027 & 40.677 & 61.616 & 398 \\
UPK 305 & 41.127 & 39.027 & 78 \\
UPK 296 & 42.666 & 48.561 & 46 \\
King 5 & 48.682 & 52.695 & 92 \\
Stock 23 & 49.075 & 60.377 & 125 \\
COIN-Gaia 38 & 51.472 & 51.072 & 73 \\
ASCC 10 & 51.87 & 34.981 & 143 \\
King 6 & 51.982 & 56.444 & 256 \\
Czernik 16 & 52.754 & 52.612 & 76 \\
NGC 1342 & 52.894 & 37.38 & 405 \\
ASCC 11 & 53.056 & 44.856 & 285 \\
Berkeley 9 & 53.167 & 52.649 & 46 \\
Berkeley 10 & 54.894 & 66.486 & 119 \\
Melotte 22 & 56.601 & 24.114 & 1227 \\
Tombaugh 5 & 56.984 & 59.07 & 418 \\
UBC 53 & 59.821 & 47.395 & 40 \\
UBC 49 & 60.217 & 59.198 & 40 \\
NGC 1496 & 61.111 & 52.668 & 132 \\
UPK 333 & 61.771 & 44.094 & 28 \\
NGC 1513 & 62.47 & 49.504 & 258 \\
NGC 1528 & 63.878 & 51.218 & 301 \\
UBC 54 & 64.747 & 46.453 & 172 \\
NGC 1545 & 65.202 & 50.221 & 158 \\
Czernik 18 & 66.963 & 30.936 & 40 \\
Melotte 25 & 67.447 & 16.948 & 439 \\
UBC 199 & 67.449 & 25.422 & 30 \\
FSR 0728 & 67.458 & 38.496 & 85 \\
NGC 1582 & 67.985 & 43.718 & 192 \\
Gulliver 11 & 67.996 & 43.62 & 104 \\
COIN-Gaia 11 & 68.11 & 39.479 & 113 \\
Berkeley 67 & 69.472 & 50.755 & 64 \\
Platais 3 & 69.976 & 71.28 & 138 \\
NGC 1647 & 71.481 & 19.079 & 572 \\
Alessi 2 & 71.602 & 55.199 & 202 \\
NGC 1662 & 72.198 & 10.882 & 249 \\
NGC 1664 & 72.763 & 43.676 & 281 \\
Czernik 19 & 74.258 & 28.763 & 73 \\
UBC 61 & 75.067 & 36.23 & 47 \\
RSG 1 & 75.508 & 37.475 & 177 \\
NGC 1708 & 75.871 & 52.851 & 134 \\
NGC 1750 & 75.926 & 23.695 & 378 \\
COIN-Gaia 15 & 76.09 & 35.831 & 154 \\
NGC 1758 & 76.175 & 23.813 & 192 \\
UBC 200 & 76.912 & 17.585 & 40 \\
NGC 1778 & 77.033 & 37.02 & 178 \\
NGC 1817 & 78.139 & 16.696 & 341 \\
COIN-Gaia 12 & 79.209 & 41.708 & 83 \\
NGC 1901 & 79.561 & -68.294 & 131 \\
COIN-Gaia 17 & 81.244 & 37.558 & 98 \\
NGC 1907 & 82.033 & 35.33 & 244 \\
NGC 1912 & 82.167 & 35.824 & 702 \\
COIN-Gaia 19 & 82.188 & 34.29 & 113 \\
UBC 59 & 82.239 & 48.043 & 32 \\
COIN-Gaia 13 & 83.186 & 42.087 & 379 \\
COIN-Gaia 26 & 83.771 & 15.721 & 105 \\
Koposov 36 & 84.218 & 31.21 & 50 \\
UBC 8 & 84.519 & 57.124 & 189 \\
COIN-Gaia 27 & 85.76 & 13.743 & 58 \\
FSR 0850 & 86.257 & 24.74 & 134 \\
Collinder 74 & 87.17 & 7.374 & 80 \\
COIN-Gaia 23 & 87.449 & 27.008 & 120 \\
NGC 2099 & 88.074 & 32.545 & 1198 \\
NGC 2112 & 88.452 & 0.403 & 568 \\
Basel 11b & 89.547 & 21.965 & 73 \\
COIN-Gaia 41 & 89.835 & 19.026 & 82 \\
Koposov 12 & 90.245 & 35.287 & 152 \\
NGC 2126 & 90.658 & 49.883 & 129 \\
UBC 72 & 90.986 & 26.651 & 81 \\
COIN-Gaia 22 & 91.06 & 31.602 & 79 \\
FSR 0932 & 91.087 & 14.573 & 27 \\
NGC 2184 & 91.69 & -2.0 & 99 \\
COIN-Gaia 25 & 91.691 & 20.276 & 131 \\
NGC 2168 & 92.272 & 24.336 & 1014 \\
FSR 0923 & 92.644 & 16.971 & 62 \\
FSR 0985 & 92.953 & 7.02 & 66 \\
NGC 2186 & 93.031 & 5.453 & 100 \\
Czernik 25 & 93.432 & 6.953 & 132 \\
FSR 0893 & 93.454 & 21.608 & 74 \\
Ferrero 11 & 93.646 & 0.637 & 109 \\
ASCC 23 & 95.047 & 46.71 & 166 \\
NGC 2215 & 95.199 & -7.277 & 112 \\
LP 658 & 95.274 & -3.431 & 119 \\
UBC 73 & 95.381 & 26.913 & 74 \\
UBC 74 & 95.461 & 22.419 & 73 \\
UPK 381 & 95.506 & 29.107 & 101 \\
FSR 0951 & 95.573 & 14.65 & 184 \\
UBC 90 & 97.208 & 14.921 & 71 \\
NGC 2236 & 97.416 & 6.834 & 272 \\
UPK 442 & 98.4 & -11.549 & 84 \\
NGC 2251 & 98.68 & 8.337 & 113 \\

        \hline
    \end{tabular}

        \begin{tabular}{p{1.8cm}ccc}
        \hline
        Cluster name & $\alpha\ /\ \mathrm{deg}$ & $\delta\ /\ \mathrm{deg}$ & $N_{P>0.9}$ \\
        \hline
    
NGC 2254 & 98.949 & 7.671 & 97 \\
Ruprecht 1 & 99.084 & -14.157 & 170 \\
Collinder 110 & 99.677 & 2.069 & 656 \\
UBC 215 & 100.461 & -5.243 & 122 \\
UBC 212 & 100.957 & -0.886 & 70 \\
NGC 2287 & 101.499 & -20.716 & 614 \\
UPK 350 & 101.689 & 48.688 & 93 \\
NGC 2286 & 101.916 & -3.167 & 138 \\
NGC 2281 & 102.091 & 41.06 & 351 \\
LP 930 & 102.845 & -1.801 & 175 \\
NGC 2301 & 102.943 & 0.465 & 544 \\
NGC 2302 & 102.977 & -7.086 & 105 \\
UPK 433 & 103.455 & -5.729 & 99 \\
NGC 2309 & 104.02 & -7.18 & 212 \\
ASCC 30 & 104.271 & -6.23 & 110 \\
NGC 2318 & 104.854 & -13.822 & 150 \\
Gulliver 13 & 104.858 & -13.254 & 44 \\
Tombaugh 1 & 105.126 & -20.569 & 215 \\
UPK 524 & 105.304 & -54.432 & 93 \\
Ruprecht 8 & 105.424 & -13.539 & 80 \\
Alessi 60 & 105.619 & -1.12 & 86 \\
NGC 2323 & 105.684 & -8.365 & 560 \\
UBC 447 & 106.389 & -9.026 & 76 \\
NGC 2335 & 106.69 & -10.023 & 128 \\
Gulliver 21 & 106.961 & -25.462 & 188 \\
NGC 2345 & 107.075 & -13.199 & 335 \\
LP 2198 & 107.506 & 4.667 & 121 \\
Gulliver 47 & 107.932 & 0.827 & 92 \\
UPK 431 & 108.234 & -2.635 & 96 \\
NGC 2354 & 108.503 & -25.724 & 230 \\
NGC 2353 & 108.641 & -10.257 & 206 \\
UPK 429 & 108.961 & -0.875 & 146 \\
NGC 2355 & 109.247 & 13.772 & 214 \\
NGC 2358 & 109.261 & -17.143 & 120 \\
Alessi 3 & 109.275 & -46.142 & 145 \\
NGC 2360 & 109.443 & -15.631 & 577 \\
Haffner 5 & 109.493 & -22.66 & 279 \\
NGC 2374 & 111.016 & -13.235 & 143 \\
Ruprecht 18 & 111.16 & -26.213 & 207 \\
Ruprecht 23 & 112.665 & -23.383 & 86 \\
UBC 233 & 112.783 & -28.377 & 81 \\
NGC 2422 & 114.147 & -14.489 & 595 \\
UBC 221 & 114.297 & -15.681 & 82 \\
NGC 2423 & 114.299 & -13.863 & 400 \\
Melotte 71 & 114.383 & -12.065 & 506 \\
Ruprecht 27 & 114.423 & -26.516 & 132 \\
NGC 2420 & 114.602 & 21.575 & 144 \\
Melotte 72 & 114.618 & -10.698 & 202 \\
NGC 2428 & 114.831 & -16.55 & 152 \\
NGC 2432 & 115.223 & -19.078 & 96 \\
NGC 2437 & 115.445 & -14.844 & 1349 \\
ESO 368 11 & 116.086 & -34.628 & 84 \\
NGC 2447 & 116.141 & -23.853 & 603 \\
Ruprecht 33 & 116.445 & -21.915 & 117 \\
ASCC 41 & 116.674 & 0.137 & 147 \\
FSR 1361 & 117.656 & -36.401 & 224 \\
NGC 2477 & 118.046 & -38.537 & 1708 \\
Alessi Teutsch 3 & 118.228 & -53.022 & 167 \\
NGC 2479 & 118.762 & -17.732 & 152 \\
NGC 2482 & 118.787 & -24.263 & 145 \\
NGC 2489 & 119.064 & -30.06 & 257 \\
UBC 229 & 119.27 & -22.796 & 106 \\
NGC 2516 & 119.527 & -60.8 & 1663 \\
NGC 2509 & 120.201 & -19.056 & 231 \\
FSR 1378 & 120.304 & -40.676 & 239 \\
UPK 467 & 121.237 & -11.42 & 127 \\
NGC 2527 & 121.246 & -28.122 & 305 \\
NGC 2539 & 122.658 & -12.834 & 340 \\
NGC 2546 & 123.082 & -37.661 & 353 \\
Gulliver 36 & 123.185 & -35.111 & 34 \\
NGC 2548 & 123.412 & -5.726 & 415 \\
UBC 239 & 123.763 & -30.155 & 97 \\
NGC 2567 & 124.645 & -30.631 & 194 \\
UPK 537 & 126.57 & -52.113 & 78 \\
Gulliver 44 & 127.249 & -38.095 & 159 \\
LP 2220 & 127.45 & -41.343 & 150 \\
Pismis 3 & 127.834 & -38.64 & 547 \\
UBC 255 & 128.456 & -61.463 & 69 \\
Pismis 4 & 128.79 & -44.407 & 227 \\
NGC 2627 & 129.309 & -29.952 & 287 \\
NGC 2632 & 130.054 & 19.621 & 794 \\
Ruprecht 67 & 130.464 & -43.363 & 82 \\
UBC 246 & 130.734 & -44.932 & 288 \\
LP 2219 & 130.848 & -45.788 & 158 \\
NGC 2670 & 131.386 & -48.801 & 209 \\
NGC 2671 & 131.557 & -41.886 & 317 \\
NGC 2669 & 131.611 & -52.931 & 229 \\
NGC 2682 & 132.846 & 11.814 & 734 \\
UBC 243 & 133.864 & -40.746 & 91 \\
UPK 508 & 135.775 & -34.942 & 67 \\
UPK 528 & 137.032 & -43.346 & 77 \\
Pismis 12 & 140.007 & -45.131 & 177 \\
Gulliver 57 & 141.203 & -48.075 & 69 \\
IC 2488 & 141.857 & -57.004 & 448 \\
NGC 2925 & 143.321 & -53.413 & 202 \\
Pismis 15 & 143.684 & -48.04 & 40 \\
NGC 2972 & 145.058 & -50.326 & 123 \\
Ruprecht 82 & 146.421 & -54.004 & 96 \\
UPK 549 & 146.49 & -52.49 & 200 \\
NGC 3033 & 147.145 & -56.415 & 79 \\
Ruprecht 84 & 147.281 & -65.259 & 81 \\
NGC 3114 & 150.553 & -60.041 & 1068 \\
BH 87 & 151.148 & -55.386 & 197 \\
Ruprecht 161 & 152.4 & -61.259 & 106 \\
LP 2059 & 153.233 & -60.885 & 73 \\
LP 5 & 156.138 & -72.51 & 199 \\
Collinder 220 & 156.429 & -57.922 & 146 \\
UBC 259 & 157.621 & -60.078 & 145 \\
UPK 560 & 158.037 & -44.451 & 65 \\
Melotte 101 & 160.535 & -65.11 & 367 \\
UPK 567 & 161.029 & -67.524 & 92 \\
Gulliver 52 & 161.669 & -59.508 & 52 \\
Ruprecht 91 & 161.964 & -57.483 & 245 \\
NGC 3496 & 164.877 & -60.335 & 265 \\
Ruprecht 93 & 166.037 & -61.372 & 160 \\
NGC 3532 & 166.417 & -58.707 & 1751 \\
LP 2238 & 167.713 & -56.821 & 87 \\
Trumpler 19 & 168.623 & -57.563 & 397 \\
IC 2714 & 169.373 & -62.719 & 443 \\

        \hline
    \end{tabular}

\begin{tabular}{p{1.8cm}ccc}
        \hline
        Cluster name & $\alpha\ /\ \mathrm{deg}$ & $\delta\ /\ \mathrm{deg}$ & $N_{P>0.9}$ \\
        \hline
    
UBC 270 & 169.813 & -59.552 & 53 \\
Melotte 105 & 169.92 & -63.486 & 295 \\
LP 1540 & 170.219 & -59.302 & 75 \\
NGC 3680 & 171.392 & -43.24 & 62 \\
UBC 277 & 174.042 & -59.46 & 148 \\
LP 2094 & 175.495 & -61.961 & 136 \\
NGC 3960 & 177.644 & -55.679 & 145 \\
Ruprecht 98 & 179.715 & -64.581 & 147 \\
Alessi Teutsch 8 & 180.649 & -60.935 & 385 \\
ESO 130 06 & 181.97 & -59.31 & 128 \\
ESO 130 08 & 182.619 & -59.479 & 214 \\
UBC 284 & 184.283 & -62.952 & 112 \\
ASCC 71 & 185.033 & -67.509 & 145 \\
Melotte 111 & 186.014 & 25.652 & 174 \\
NGC 4337 & 186.022 & -58.125 & 183 \\
NGC 4349 & 186.048 & -61.866 & 385 \\
BH 132 & 186.72 & -64.058 & 86 \\
ASCC 73 & 189.281 & -67.203 & 51 \\
LP 1994 & 191.005 & -58.096 & 141 \\
Gulliver 58 & 191.515 & -61.965 & 115 \\
UBC 290 & 191.949 & -59.381 & 201 \\
UPK 579 & 192.917 & -50.541 & 140 \\
UPK 578 & 192.965 & -57.367 & 54 \\
Collinder 268 & 199.543 & -67.079 & 115 \\
Collinder 269 & 200.992 & -66.204 & 58 \\
UBC 294 & 204.258 & -58.073 & 38 \\
NGC 5269 & 206.147 & -62.907 & 100 \\
Collinder 277 & 207.058 & -66.066 & 157 \\
NGC 5288 & 207.193 & -64.68 & 52 \\
UBC 295 & 207.554 & -60.399 & 49 \\
NGC 5316 & 208.516 & -61.883 & 334 \\
NGC 5381 & 210.205 & -59.578 & 300 \\
NGC 5460 & 211.847 & -48.285 & 180 \\
Loden 1194 & 212.006 & -59.786 & 81 \\
Ruprecht 167 & 215.15 & -58.858 & 84 \\
Lynga 2 & 216.085 & -61.328 & 158 \\
NGC 5617 & 217.452 & -60.719 & 311 \\
NGC 5662 & 218.734 & -56.64 & 334 \\
Ruprecht 111 & 219.01 & -59.975 & 102 \\
Alessi 6 & 220.058 & -66.127 & 146 \\
NGC 5715 & 220.859 & -57.578 & 95 \\
UBC 303 & 221.85 & -56.255 & 34 \\
UBC 300 & 222.958 & -65.308 & 48 \\
Teutsch 80 & 223.361 & -60.478 & 49 \\
NGC 5822 & 226.051 & -54.366 & 545 \\
LP 2100 & 227.612 & -60.495 & 52 \\
NGC 5925 & 231.847 & -54.515 & 197 \\
Alessi 8 & 232.408 & -51.227 & 103 \\
UFMG 1 & 236.593 & -56.792 & 141 \\
Collinder 292 & 237.41 & -57.594 & 81 \\
UFMG 2 & 237.585 & -55.961 & 187 \\
UFMG 3 & 238.115 & -55.419 & 119 \\
NGC 6005 & 238.955 & -57.439 & 117 \\
Trumpler 23 & 240.218 & -53.539 & 116 \\
NGC 6025 & 240.779 & -60.43 & 394 \\
Ruprecht 115 & 243.215 & -52.393 & 49 \\
NGC 6067 & 243.299 & -54.227 & 577 \\
UBC 314 & 243.386 & -50.128 & 77 \\
Ruprecht 176 & 243.676 & -51.332 & 105 \\
UBC 542 & 244.575 & -49.487 & 56 \\
NGC 6087 & 244.721 & -57.916 & 284 \\
Harvard 10 & 244.859 & -54.957 & 257 \\
Lynga 9 & 245.17 & -48.523 & 153 \\
NGC 6124 & 246.332 & -40.661 & 819 \\
UBC 315 & 246.376 & -47.935 & 52 \\
NGC 6134 & 246.953 & -49.161 & 392 \\
NGC 6152 & 248.213 & -52.629 & 181 \\
NGC 6167 & 248.662 & -49.759 & 750 \\
Collinder 307 & 248.792 & -50.99 & 49 \\
UPK 644 & 249.155 & -31.418 & 86 \\
UBC 319 & 249.699 & -44.755 & 121 \\
NGC 6192 & 250.077 & -43.355 & 401 \\
Ruprecht 121 & 250.436 & -46.159 & 388 \\
ESO 518 03 & 251.76 & -25.81 & 6 \\
UBC 669 & 251.81 & -46.665 & 36 \\
UBC 321 & 251.906 & -44.536 & 51 \\
NGC 6208 & 252.336 & -53.714 & 222 \\
BH 200 & 252.484 & -44.182 & 171 \\
UPK 612 & 253.423 & -67.788 & 226 \\
BH 202 & 253.779 & -40.947 & 161 \\
UBC 671 & 253.992 & -43.207 & 41 \\
NGC 6253 & 254.778 & -52.712 & 218 \\
NGC 6259 & 255.195 & -44.678 & 443 \\
UBC 322 & 255.213 & -44.184 & 85 \\
NGC 6268 & 255.524 & -39.721 & 152 \\
BH 211 & 255.535 & -41.113 & 61 \\
UBC 324 & 255.954 & -38.465 & 0 \\
Teutsch 84 & 256.09 & -42.07 & 40 \\
NGC 6281 & 256.179 & -37.948 & 433 \\
ASCC 88 & 256.886 & -35.564 & 138 \\
LP 145 & 256.976 & -44.157 & 33 \\
UBC 326 & 258.155 & -36.056 & 44 \\
Trumpler 25 & 261.125 & -39.006 & 402 \\
IC 4651 & 261.212 & -49.917 & 558 \\
LP 866 & 261.768 & -39.405 & 98 \\
Trumpler 26 & 262.126 & -29.487 & 122 \\
Collinder 338 & 264.483 & -37.657 & 243 \\
ASCC 90 & 264.777 & -34.874 & 127 \\
NGC 6404 & 264.916 & -33.224 & 345 \\
NGC 6400 & 265.062 & -36.957 & 129 \\
Alessi 9 & 265.974 & -47.028 & 269 \\
NGC 6416 & 266.014 & -32.344 & 257 \\
Ruprecht 128 & 266.063 & -34.879 & 101 \\
LP 1624 & 266.582 & -29.175 & 268 \\
NGC 6425 & 266.733 & -31.507 & 177 \\
Collinder 350 & 267.018 & 1.525 & 180 \\
Alessi 31 & 267.698 & -11.853 & 188 \\
Ruprecht 134 & 268.184 & -29.537 & 65 \\
UBC 95 & 268.252 & -22.171 & 69 \\
Czernik 37 & 268.32 & -27.373 & 157 \\
NGC 6475 & 268.447 & -34.841 & 1133 \\
Trumpler 30 & 269.182 & -35.298 & 127 \\
NGC 6494 & 269.237 & -18.987 & 760 \\
Ruprecht 135 & 269.511 & -11.646 & 106 \\
UBC 92 & 269.883 & -26.655 & 151 \\
UPK 645 & 270.167 & -42.783 & 100 \\
NGC 6568 & 273.192 & -21.612 & 304 \\
UPK 5 & 273.304 & -18.308 & 127 \\
Gulliver 20 & 273.736 & 11.082 & 91 \\

        \hline
    \end{tabular}

    \begin{tabular}{p{1.8cm}ccc}
        \hline
        Cluster name & $\alpha\ /\ \mathrm{deg}$ & $\delta\ /\ \mathrm{deg}$ & $N_{P>0.9}$ \\
        \hline
    
NGC 6583 & 273.962 & -22.143 & 183 \\
LP 1218 & 273.976 & -18.099 & 82 \\
Trumpler 32 & 274.294 & -13.349 & 311 \\
UPK 27 & 275.051 & -5.167 & 73 \\
Ferrero 1 & 275.063 & -32.338 & 121 \\
Mamajek 4 & 276.635 & -50.772 & 217 \\
NGC 6633 & 276.845 & 6.615 & 353 \\
IC 4725 & 277.937 & -19.114 & 557 \\
Ruprecht 171 & 278.012 & -16.062 & 486 \\
NGC 6645 & 278.158 & -16.918 & 216 \\
UPK 29 & 278.617 & -5.47 & 92 \\
NGC 6664 & 279.125 & -8.194 & 205 \\
UBC 32 & 279.312 & -14.096 & 228 \\
IC 4756 & 279.649 & 5.435 & 431 \\
UBC 106 & 280.475 & -5.417 & 462 \\
UPK 20 & 281.145 & -14.456 & 99 \\
Berkeley 79 & 281.25 & -1.146 & 153 \\
NGC 6694 & 281.317 & -9.386 & 199 \\
UPK 4 & 281.361 & -23.764 & 115 \\
UBC 108 & 281.611 & -5.181 & 66 \\
LP 1235 & 281.846 & -5.255 & 175 \\
UPK 13 & 282.007 & -18.281 & 136 \\
Basel 1 & 282.036 & -5.876 & 94 \\
LP 597 & 282.104 & 8.714 & 85 \\
ASCC 99 & 282.17 & -18.488 & 188 \\
LP 2117 & 282.302 & -4.466 & 524 \\
Czernik 38 & 282.451 & 4.965 & 126 \\
Ruprecht 145 & 282.636 & -18.27 & 0 \\
NGC 6704 & 282.687 & -5.203 & 139 \\
NGC 6705 & 282.766 & -6.272 & 848 \\
NGC 6709 & 282.836 & 10.334 & 321 \\
UPK 21 & 282.895 & -15.15 & 88 \\
UPK 50 & 283.309 & 8.172 & 129 \\
Berkeley 80 & 283.591 & -1.214 & 81 \\
LP 2068 & 283.967 & -5.094 & 138 \\
Alessi 62 & 284.026 & 21.597 & 173 \\
NGC 6728 & 284.715 & -8.953 & 142 \\
NGC 6735 & 285.155 & -0.498 & 188 \\
UPK 54 & 286.157 & 15.689 & 155 \\
NGC 6755 & 286.942 & 4.224 & 146 \\
ASCC 101 & 288.399 & 36.369 & 90 \\
UPK 53 & 288.544 & 14.392 & 161 \\
UPK 12 & 288.805 & -22.143 & 109 \\
Ruprecht 147 & 289.087 & -16.333 & 179 \\
UBC 119 & 290.628 & 8.41 & 17 \\
NGC 6793 & 290.817 & 22.159 & 206 \\
UPK 45 & 291.696 & 0.212 & 210 \\
Gulliver 37 & 292.077 & 25.347 & 87 \\
King 26 & 292.237 & 14.882 & 72 \\
Gulliver 28 & 293.559 & 18.059 & 73 \\
Teutsch 35 & 294.091 & 35.742 & 203 \\
Stock 1 & 294.146 & 25.163 & 182 \\
UBC 123 & 294.181 & 18.247 & 61 \\
NGC 6811 & 294.34 & 46.378 & 251 \\
Alessi 44 & 295.325 & 1.592 & 117 \\
NGC 6819 & 295.327 & 40.19 & 831 \\
Skiff J1942+38.6 & 295.611 & 38.645 & 36 \\
LP 321 & 296.389 & 21.153 & 21 \\
UPK 55 & 296.812 & 10.428 & 92 \\
NGC 6830 & 297.718 & 23.101 & 268 \\
ASCC 108 & 298.306 & 39.349 & 250 \\
UPK 93 & 300.324 & 29.971 & 52 \\
ASCC 110 & 300.742 & 33.528 & 77 \\
FSR 0241 & 300.763 & 46.196 & 48 \\
Gulliver 38 & 300.808 & 34.435 & 92 \\
NGC 6866 & 300.983 & 44.158 & 207 \\
Gulliver 31 & 301.912 & 38.232 & 93 \\
ASCC 111 & 302.891 & 37.515 & 156 \\
Alessi Teutsch 11 & 304.127 & 52.051 & 176 \\
UBC 136 & 304.49 & 32.582 & 163 \\
Collinder 421 & 305.829 & 41.701 & 147 \\
FSR 0278 & 307.761 & 51.021 & 66 \\
NGC 6939 & 307.917 & 60.653 & 184 \\
NGC 6940 & 308.626 & 28.278 & 371 \\
UPK 84 & 310.245 & 20.212 & 128 \\
Ruprecht 174 & 310.856 & 37.031 & 95 \\
Alessi 12 & 310.875 & 23.871 & 284 \\
Roslund 7 & 313.196 & 37.847 & 216 \\
Barkhatova 1 & 313.398 & 46.037 & 119 \\
NGC 6991 & 313.621 & 47.4 & 148 \\
Gulliver 30 & 313.673 & 45.996 & 80 \\
UPK 51 & 314.073 & -6.588 & 104 \\
NGC 6997 & 314.128 & 44.64 & 289 \\
NGC 7031 & 316.809 & 50.87 & 168 \\
ASCC 113 & 317.933 & 38.638 & 298 \\
UBC 155 & 318.413 & 50.134 & 0 \\
UPK 143 & 319.134 & 49.942 & 120 \\
Teutsch 144 & 320.438 & 50.595 & 76 \\
NGC 7062 & 320.862 & 46.385 & 100 \\
UBC 156 & 321.715 & 49.114 & 58 \\
FSR 0306 & 322.129 & 51.69 & 68 \\
NGC 7086 & 322.624 & 51.593 & 490 \\
UPK 136 & 322.841 & 43.441 & 82 \\
NGC 7092 & 322.889 & 48.247 & 237 \\
NGC 7142 & 326.29 & 65.782 & 156 \\
NGC 7209 & 331.224 & 46.508 & 297 \\
FSR 0342 & 331.931 & 53.104 & 260 \\
NGC 7243 & 333.788 & 49.83 & 329 \\
UPK 194 & 336.446 & 65.004 & 137 \\
NGC 7296 & 337.0 & 52.321 & 123 \\
UPK 167 & 337.143 & 49.814 & 126 \\
LP 1800 & 337.843 & 58.053 & 214 \\
UPK 180 & 339.051 & 56.771 & 113 \\
FSR 0385 & 340.628 & 58.941 & 121 \\
Alessi 37 & 341.961 & 46.342 & 266 \\
FSR 0384 & 342.911 & 56.109 & 38 \\
UBC 172 & 343.401 & 54.336 & 85 \\
UBC 6 & 344.01 & 51.187 & 160 \\
UBC 175 & 344.451 & 54.401 & 121 \\
King 19 & 347.053 & 60.523 & 126 \\
ASCC 128 & 349.949 & 54.435 & 202 \\
Gulliver 49 & 350.704 & 61.988 & 154 \\
UPK 220 & 350.955 & 66.505 & 218 \\
NGC 7654 & 351.195 & 61.59 & 755 \\
FSR 0442 & 352.489 & 63.447 & 135 \\
Stock 12 & 353.923 & 52.685 & 178 \\
UBC 180 & 357.351 & 59.002 & 35 \\
NGC 7762 & 357.472 & 68.035 & 342 \\
NGC 7789 & 359.334 & 56.726 & 1596 \\

        \hline
    \end{tabular}
    
\end{multicols*}
\end{minipage}
    \normalsize
\end{table*}

\endgroup

\section{Data Availability}

The data used in this paper are available from the \textit{Gaia} DR3 science 
archive, at \url{gea.esac.esa.int}. Queries used to duplicate the dataset used in this work are described in Appendix \ref{sec:query}. Data described in Tables \ref{tab:clusters} and \ref{tab:members} (i. e. parameters of analysed clusters and a list of cluster members) are available as supplementary material to this paper and at \url{https://vizier.cds.unistra.fr/viz-bin/VizieR}. Simulations of cluster dissolution will be shared via private communication with a reasonable request.

\section{Gaia queries}
\label{sec:query}

The volume we analysed was divided into sections. We conducted separate \textit{Gaia} data queries for each section. The \texttt{ADQL} query for one section is:

\begin{spverbatim}
SELECT gaiadr3.gaia_source.source_id, gaiadr3.gaia_source.ra, gaiadr3.gaia_source.dec, gaiadr3.gaia_source.pmra, gaiadr3.gaia_source.pmdec, gaiadr3.gaia_source.parallax, gaiadr3.gaia_source.radial_velocity, gaiadr3.gaia_source.phot_g_mean_mag, gaiadr3.gaia_source.phot_rp_mean_mag, gaiadr3.gaia_source.phot_bp_mean_mag, external.gaiaedr3_distance.r_med_geo, gaiadr3.astrophysical_parameters.teff_gspspec, gaiadr3.astrophysical_parameters.logg_gspspec, gaiadr3.astrophysical_parameters.mh_gspspec, gaiadr3.astrophysical_parameters.alphafe_gspspec, gaiadr3.gaia_source.ag_gspphot, gaiadr3.gaia_source.ebpminrp_gspphot, gaiadr3.gaia_source.ruwe 
FROM gaiadr3.gaia_source 
LEFT JOIN external.gaiaedr3_distance 
ON gaiadr3.gaia_source.source_id = external.gaiaedr3_distance.source_id 
LEFT JOIN gaiadr3.astrophysical_parameters 
ON gaiadr3.gaia_source.source_id = gaiadr3.astrophysical_parameters.source_id 
WHERE gaiadr3.gaia_source.ra >= 94.73684210526315 
AND gaiadr3.gaia_source.ra < 113.68421052631578 
AND gaiadr3.gaia_source.dec >= -33.15789473684211 
AND gaiadr3.gaia_source.dec < -23.684210526315795 
AND gaiadr3.gaia_source.parallax >= 0.36291930060041844 
AND gaiadr3.gaia_source.parallax < 0.3806079094531518 
AND gaiadr3.gaia_source.phot_g_mean_mag < 17.5 
AND gaiadr3.gaia_source.phot_bp_mean_mag < 20 
AND gaiadr3.gaia_source.phot_rp_mean_mag < 20
\end{spverbatim}
\vspace{1ex}

The limits for \texttt{ra}, \texttt{dec}, and \texttt{parallax} were different for each section, and the numbers above only serve as placeholders. 

\section{\textsc{Galaxia} parameters}

Below is the parameter file for one \textsc{Galaxia} simulation.

\begin{spverbatim}
outputFile galaxia_2947
modelFile Model/population_parameters_mrtd5.ebf
outputDir ./galaxia_cache_all
photoSys parsec1/GAIADR2_TMASS
magcolorNames gaia_g,gaia_gbp-gaia_grp
appMagLimits[0] -10
appMagLimits[1] 17.5
absMagLimits[0] -1000
absMagLimits[1] 1000
colorLimits[0] -1000
colorLimits[1] 1000
geometryOption 1
longitude 239.15335157491867
latitude -11.453305872563032
surveyArea 281.3196664594758
fSample 5.0
popID -1
warpFlareOn 1
seed 1
r_max 2.755433503662073
starType 0
photoError 0
\end{spverbatim}
\vspace{1ex}

Note that the \textsc{Galaxia} simulation can only be run within a cone, while Gaia queries were done within sections limited by $\alpha$, $\delta$. For each section we ran the \textsc{Galaxia} simulation within a cone covering the whole section and later discarded the stars outside that section. \textsc{Galaxia} can be run for the whole sky, but our computer did not have enough memory to run such a simulation with sufficient oversampling.

\section{\textsc{Galpy} gravitational potential of the Galaxy}
\label{sec:galpy}

Below is the code to construct the gravitational potential of the Galaxy used in this work from \textsc{galpy} potentials in \textsc{python}.

\begin{spverbatim}
from galpy import potential

RO = 8.122 #kpc

diskp = potential.MiyamotoNagaiPotential(normalize = 0.54, a = 3.0/RO, b = 0.28/RO)
halop = potential.NFWPotential(normalize = 0.35, a = 2.0)
sbp = potential.SoftenedNeedleBarPotential(normalize = 0.1, a = 3.5/RO, c = 1.0/RO, omegab = 1.85)
npot = potential.PlummerPotential(normalize = 0.01, b = 0.25/RO)
	
pot = [diskp, halop, npot, sbp]
\end{spverbatim}

\end{appendix}

\end{document}